\documentclass[aps,pra,preprint,superscriptaddress,longbibliography]{revtex4-2}

\usepackage{amsbsy,amstext,amscd,amsxtra,amsopn,mathtools}
\usepackage{tabularx}
\usepackage{amsmath}
\usepackage{amsthm}
\usepackage{amsfonts}
\usepackage{amssymb}
\usepackage{xcolor,graphicx}
\usepackage{tikz}
\usepackage{color}
\usepackage[pdftex]{hyperref} 
\usepackage{longtable}
\usepackage{array}
\usepackage{dsfont}
\usepackage{multirow}
\hypersetup{
    colorlinks=true,
    linkcolor=blue,
    filecolor=blue,      
    urlcolor=blue,
    pdftitle={EQUILATERAL TRIANGULAR WAVEGUIDES, MODAL STRUCTURE, ATTENUATION CHARACTERISTICS AND QUALITY FACTOR.},
    pdfpagemode=FullScreen,
    }
\usepackage{url}
\usepackage{balance}
\usepackage{mathrsfs}
\begin{document}
	
	\title{\textsc{EQUILATERAL TRIANGULAR WAVEGUIDES, MODAL STRUCTURE, \\ATTENUATION CHARACTERISTICS AND QUALITY FACTOR.}}
	
	\author{Francis Emenike Onah}
	\email[e-mail: ]{A00834081@tec.mx}
	\affiliation{Photonics and Mathematical Optics Group, Tecnológico de Monterrey, Mexico, 64849}
    \affiliation{The Division of Theoretical Physics, Physics and Astronomy, University of Nigeria Nsukka, Nsukka Campus, Enugu State, Nigeria.}
	
	\author{Julio C. Gutiérrez-Vega}
    \email[e-mail: ]{juliocesar@tec.mx}
	\affiliation{Photonics and Mathematical Optics Group, Tecnológico de Monterrey, Mexico, 64849}	
	
	\date{\today}
	
\begin{abstract}
    Equilateral triangular waveguides are one of the very few special kind of waveguides, whose field solutions can be constructed without necessarily solving the Maxwell's equations. Solutions can be obtained simply by superposing some plane wave solutions and requiring the solutions to obey the necessary boundary conditions for each modes, as would be expected of any Maxwell equation's solution or that of any Helmholtz equation resulting from the problem of Heat flow, membrane vibrations, etc. In fact, this was how one of the earlier solutions or eigen functions found by Gabriel Lam\'e, were obtained. Julian Schwinger, performed the same analysis, but used the superposition of complex exponential functions, in order to obtain the eigen functions of an equilateral triangular waveguide. The solutions exhibit other special symmetric properties that leave their solutions invariant and these special symmetries and applications, are what we investigate in this work, particularly as it relates to their attenuation characteristics and quality factors. Finally, by employing the number theory of Eisenstein integers or primes we have obtained a well ordered and more rigorous  eigenvalues of the equilateral triangular waveguides, which is definitely something missing in the literature. 
\end{abstract}

	
	\maketitle
	\newpage

\section{Introduction}

The mode functions, field plots, cutoff wavenumber and other properties of triangular waveguides of various cross sections have been investigated in detail by Overfelt and White\cite{Overfelt} and other researchers\cite{Isaac}, who also obtained some field solutions by the superposition of plane waves solutions\cite{Overfelt,OverfeltII}, which was applied earlier by Julian Schwinger\cite{Milton}. These solutions are not necessarily in closed form. Among all the properties of the triangular waveguides, that have been studied, no known analytic solution for the attenuation constants of the ETW has been reported until now. They are reported in this work for the first time with explicit expressions, to the best of our knowledge. 

The equilateral triangular waveguide is one peculiar waveguide structure whose plane waves superposed solutions can be written/found in closed form\cite{McCartinII}, and unlike the attenuation characteristics of other waveguides, the parabolic waveguides\cite{NoriegaM} and confocal annular elliptic waveguides\cite{GutierrezVegaIV} whose attenuation constants could only be found by numerically solving the required integrals, since the mode functions are products of  the (complicated) parabolic and Mathieu functions respectively. 

In this paper, we have used the complete solutions (superposed symmetric and anti-symmetric) of each mode to obtain completely analytic expressions of the attenuation constants and quality factors of both the E- and H-modes of the ETW. Although the quality factor expressions were given in a recent work by Mor\'an-L\'opez et al \cite{AMoranLopez}, after a "\textit{careful process}", they did not however, give the explicit expressions of the integrals around the contours, which is required for the calculation of the quality factors. Besides this latest attempt, every other calculations or evaluations of the q-factor of the ETW has been numerical, with the earlier works of Huang et al \cite{YZHuang,YZHuangII}. In another article\cite{AMoranLopezIII} Mor\'an-L\'opez et al, gave an incomplete list of the eigenvalues (\textit{the modes excludes negative integer values, and other possible $m$ and $n$ indices for each eigenvalue}) of the equilateral waveguide. As is traditionally done, we have remedied this gap\cite{Onah-PhDthesis}, by using the number theory of the Eisenstein integers or primes and the \textit{prime factorization theorem}\cite{GHHardy,DACox,KHRosen,McCartinII,McCartinIII}. 

We have given the explicit expressions for the integrals and attenuation constant and q-factor formulae, for the very first time. In both cases we have given the very well known, general expressions for the attenuation constants\cite{Borgnis}, and have derived the analogous (general) expressions for the quality factors, which can be used to obtain the quality factor of any cylindrical cavity; rectangular, circular, elliptic and parabolic cavities, once the mode functions/fields are known. We have therefore, given the complete, explicit and analytic expressions for the quality factors of the ETW, using precisely the same integral solutions, that we used to calculate the attenuation constants of the ETW.

We have also given several mode plots of the ETW solutions, to show their propagation characteristics and have also given the attenuation characteristics curves for the ETW, and as would be seen, these have the expected general properties for any waveguide, just as the rectangular\cite{Borgnis}, circular\cite{Borgnis}, elliptic\cite{GutierrezVegaIV} and parabolic\cite{GutierrezVegaIV} cylindrical waveguides. Our quality factor plots also reproduces what would be expected of the general quality factor plots in the literature. Throughout our analysis, we have used Matlab and Mathematica for our plots and verification process.

\section{Modal Structure}

\begin{figure}
\centering
\includegraphics[width=1.0\textwidth]{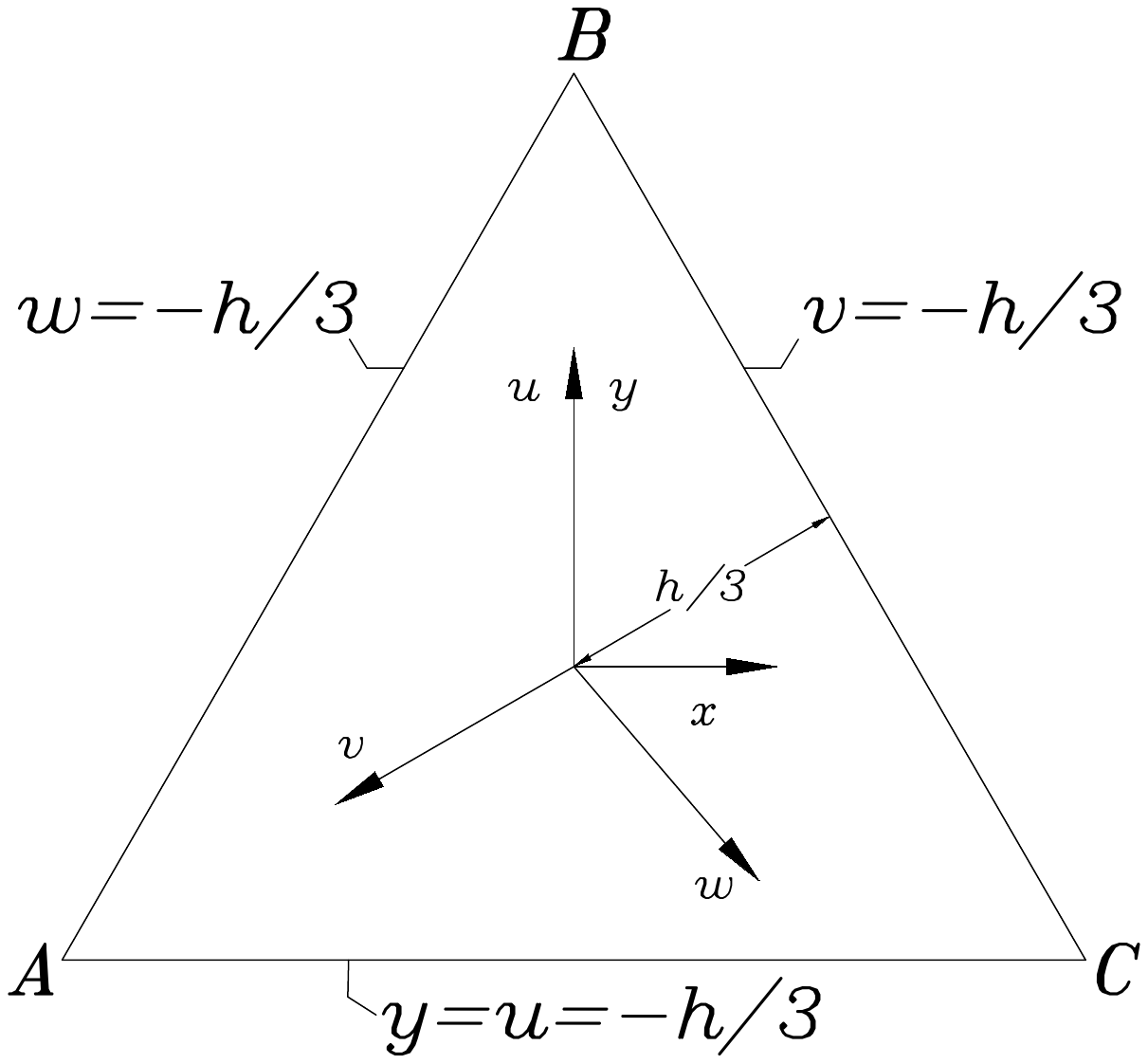}
\caption{An Equilateral Triangular Waveguide Cross Section}
\label{ETWsketch}
\end{figure}

The geometric structure of the ETW is as shown in fig:\ref{ETWsketch}.

If we defined
\begin{align}
 &Q\left(y\right)=\frac{\pi}{h}\left(y-\frac{2h}{3}\right)\quad{;}\quad\theta_{1}=\frac{\pi}{\sqrt{3}h}\left(m-n\right),\nonumber\\
 &\theta_{2}=\frac{\pi}{\sqrt{3}h}\left(n-l\right)\quad{;}\quad\theta_{3}=\frac{\pi}{\sqrt{3}h}\left(l-m\right),
 \label{mode.parameters}
\end{align}

then, we can write summarily a la Schwinger\cite{Milton,Borgnis}
\begin{align}
_{\xi}\Phi_{lmn}\left(x,y\right)=&2i\left[\sin\left(lQ\left(y\right)\right)\left(_{\xi}f\left(\theta_{1}x\right)\right)\right.\nonumber\\
&\left.+\sin\left(mQ\left(y\right)\right)\left(_{\xi}f\left(\theta_{2}x\right)\right)+\sin\left(nQ\left(y\right)\right)\left(_{\xi}f\left(\theta_{3}x\right)\right)\right],
\label{E.modes}
\end{align}

and
\begin{align}
_{\xi}\Psi_{lmn}\left(x,y\right)=&2\left[\cos\left(lQ\left(y\right)\right)\left(_{\xi}f\left(\theta_{1}x\right)\right)\right.\nonumber\\
&\left.+\cos\left(mQ\left(y\right)\right)\left(_{\xi}f\left(\theta_{2}x\right)\right)+\cos\left(nQ\left(y\right)\right)\left(_{\xi}f\left(\theta_{3}x\right)\right)\right],
\label{H.modes}
\end{align}

where
\begin{align}
_{\xi}f\left(u\right)=\begin{cases}
\cos\left(u\right), & \text{for $\xi=e$}\\
\sin\left(u\right), & \text{for $\xi=o$}.
		 \end{cases}
\end{align}

The modal structure expressions, are even or odd with respect to the x variable (the axis of symmetry is the y-axis). Using Eq.s(\ref{E.modes}) and (\ref{H.modes}) the TM fields become
\begin{align}	&E_{||}=E_{z}=\left(_{\xi}\Phi_{lmn}\left(x,y\right)\right)e^{\pm i\beta z-i\omega t},\nonumber\\
&_{\xi}\mathbf{E}_{\bot}=\pm i\frac{\beta}{k_{\bot}^{2}}\mathbf{\nabla}_{\bot}E_{||}\quad{;}\quad H_{||}=0,\nonumber\\
&_{\xi}\mathbf{H}_{\bot}=\pm\frac{1}{Z_{1}}\left(\hat{z}\times\left(_{\xi}\mathbf{E}_{\bot}\right)\right)\quad{;}\quad Z_{1}=\frac{\beta}{\omega\epsilon},
\end{align}

with $\xi$ representing the parity, in essence, for the even mode $\xi=e$ and for the odd mode $\xi=o$.

Thus, the TE fields become
\begin{align}
&H_{||}=H_{z}=\left(_{\xi}\Psi_{lmn}\left(x,y\right)\right)e^{\pm i\beta z-i\omega t},\nonumber\\
&_{\xi}\mathbf{H}_{\bot}=\pm i\frac{\beta}{k_{\bot}^{2}}\mathbf{\nabla}_{\bot}H_{||}\quad{;}\quad E_{||}=0,\nonumber\\
&_{\xi}\mathbf{E}_{\bot}=\pm Z_{2}\hat{z}\times\left(_{\xi}\mathbf{H}_{\bot}\right)\quad{;}\quad Z_{2}=\frac{\omega\mu}{\beta}.
\end{align}

Where $k_{\bot}^{2}=\gamma_{mn}^{2}=\omega^{2}\mu\epsilon-\beta^{2}$. Their respective transverse fields are given in the appendix:\ref{TMTEtransverseFields}.

\begin{figure}
\centering
\includegraphics[width=1\textwidth]{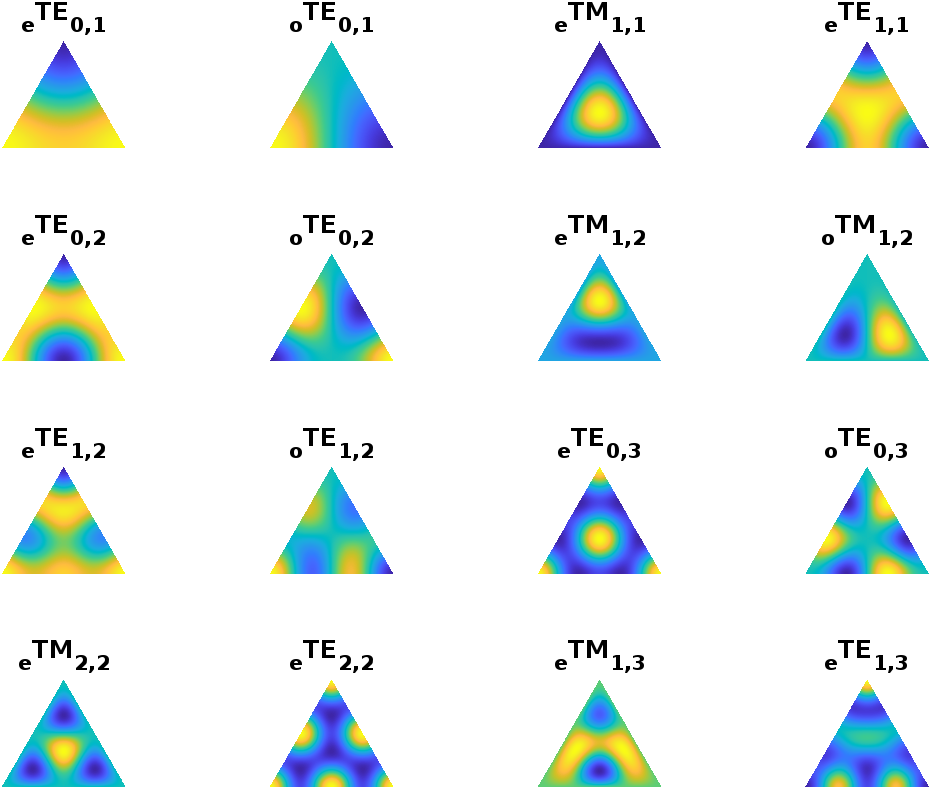}
\caption{figure}{TM and TE-Mode Structure}
\label{fig:1a}
\end{figure}

\begin{figure}[!ht]
\centering
\begin{minipage}{0.49\textwidth}
\includegraphics[width=0.56\textwidth]{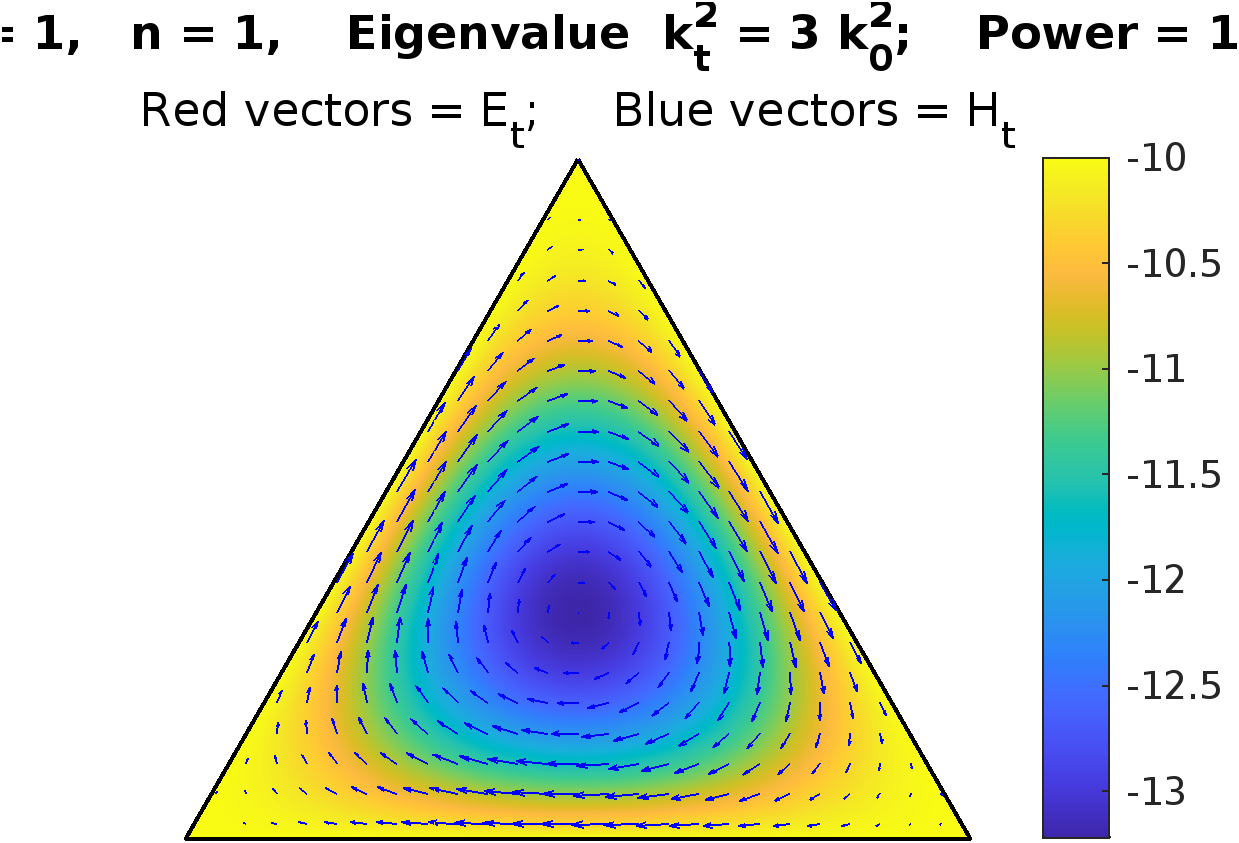}
\caption{\label{fig:Fig_a}$m=n=1$}
\end{minipage}
\hfill
\begin{minipage}{0.49\textwidth}
\includegraphics[width=0.5\textwidth]{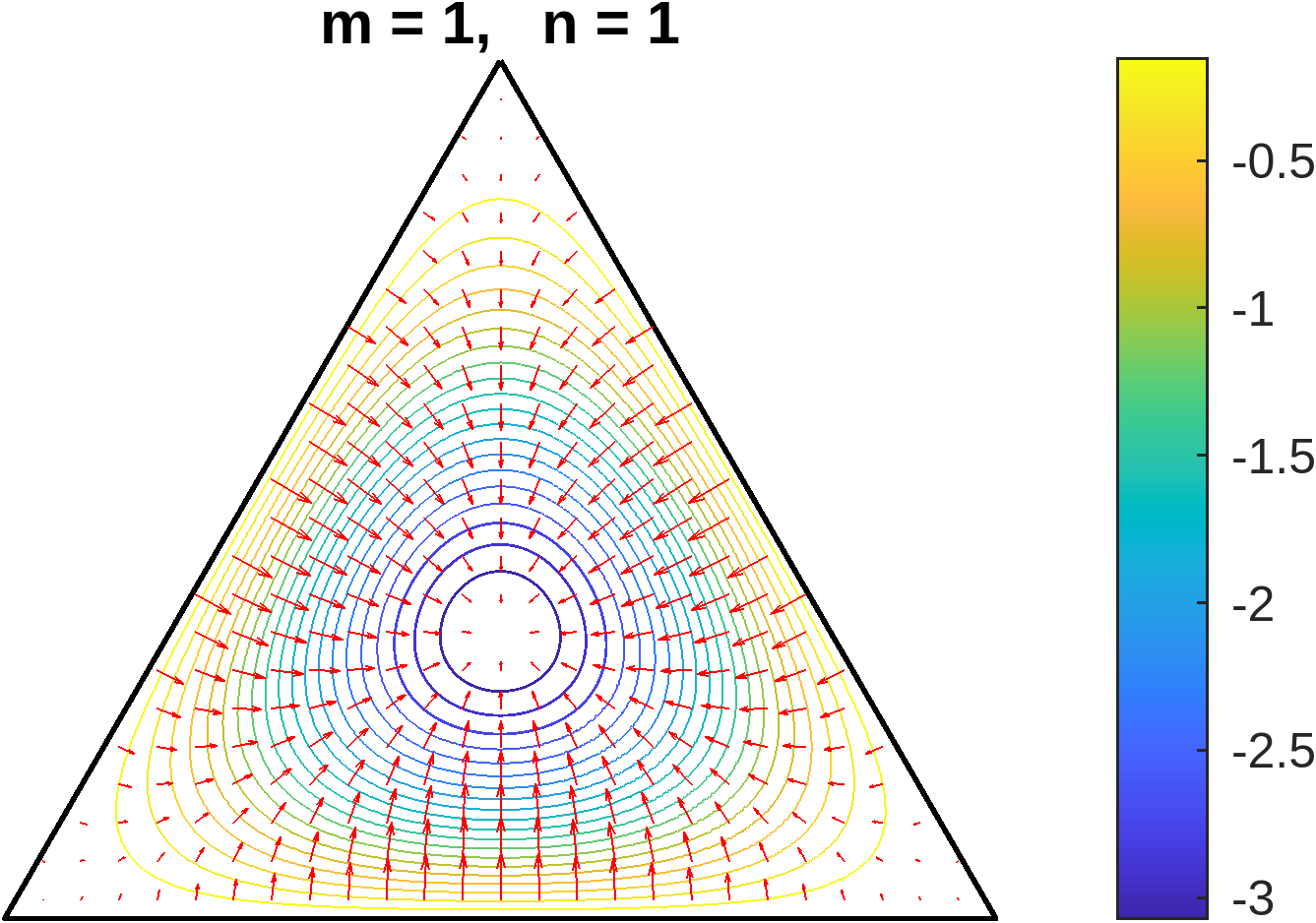}
\caption{\label{fig:Fig_b}$m=n=1$}
\end{minipage}

\begin{minipage}{0.49\textwidth}
\includegraphics[width=0.4\textwidth]{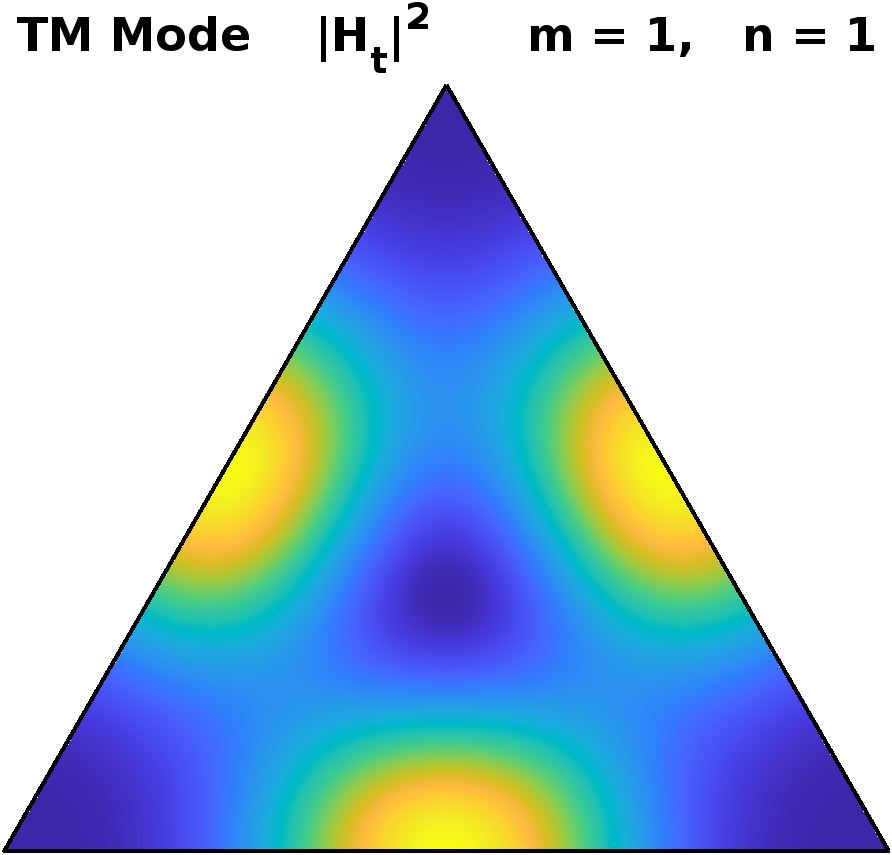}
\caption{\label{fig:Fig_c}$m=n=1$}
\end{minipage}
\hfill
\begin{minipage}{0.49\textwidth}
\includegraphics[width=0.58\textwidth]{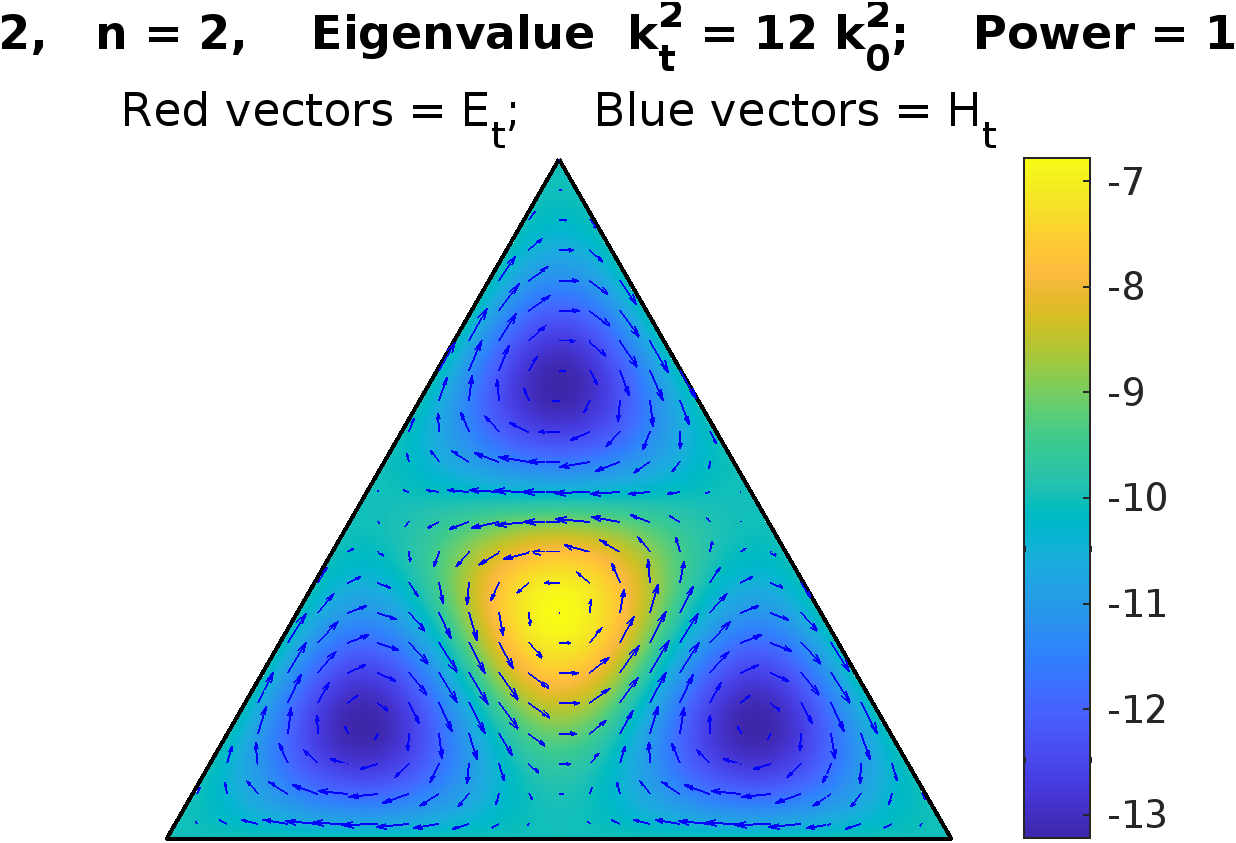}
\caption{\label{fig:Fig_d}$m=n=2$}
\end{minipage}

\begin{minipage}{0.49\textwidth}
\includegraphics[width=0.5\textwidth]{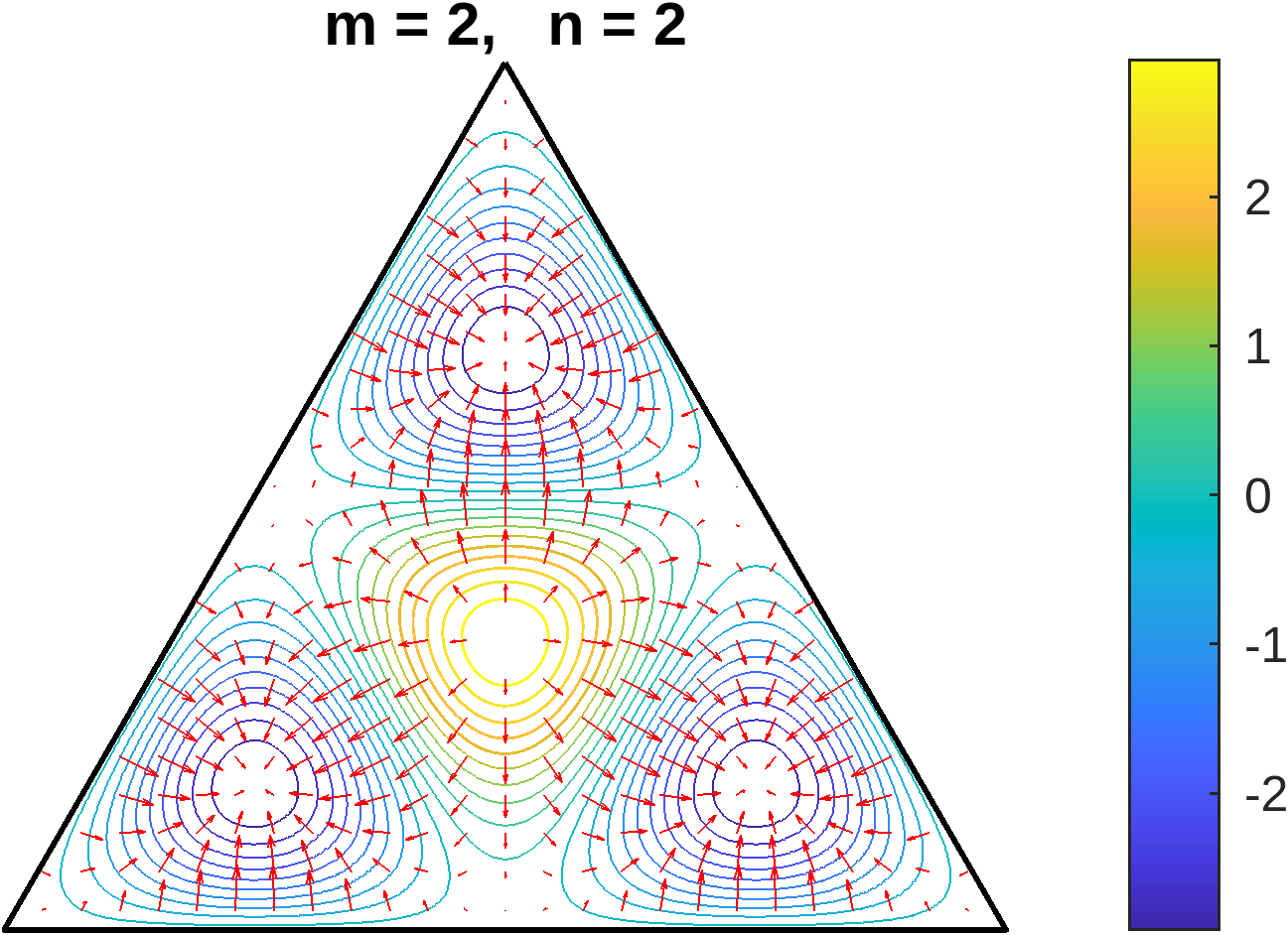}
\caption{\label{fig:Fig_e}$m=n=2$}
\end{minipage}
\hfill
\begin{minipage}{0.49\textwidth}
\includegraphics[width=0.45\textwidth]{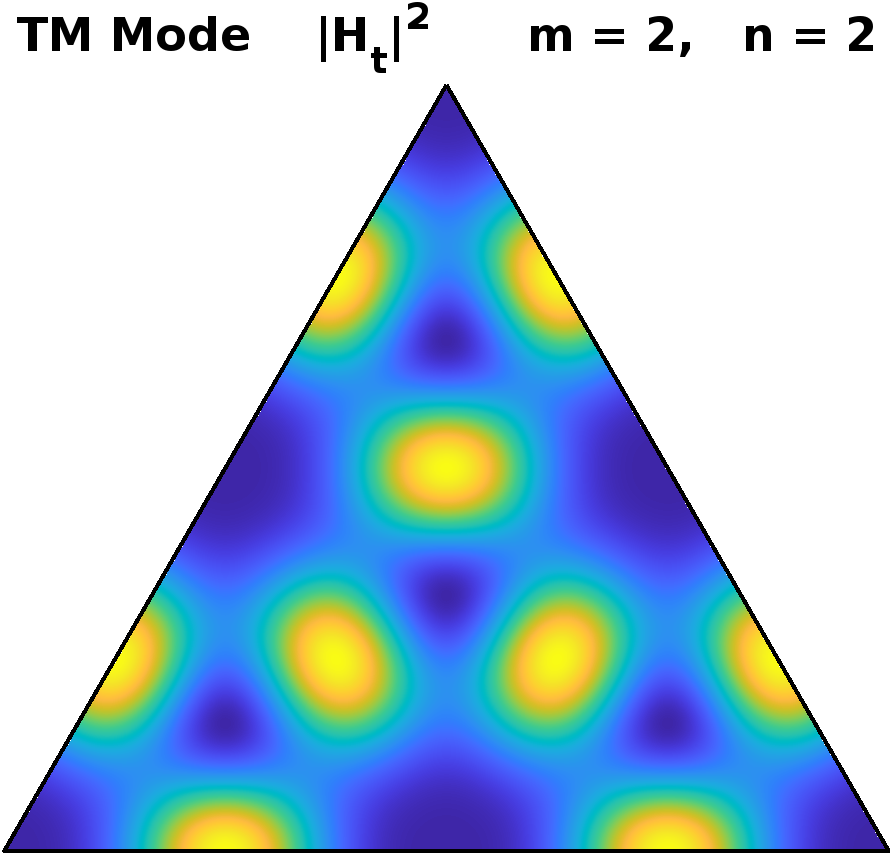}
\caption{\label{fig:Fig_f}$m=n=2$}
\end{minipage}
\caption{Transverse Fields Plots I}
\label{FigTP}
\end{figure}

\begin{figure}[!ht]
      \centering
      \begin{minipage}{0.49\textwidth}
		\includegraphics[width=0.55\linewidth]{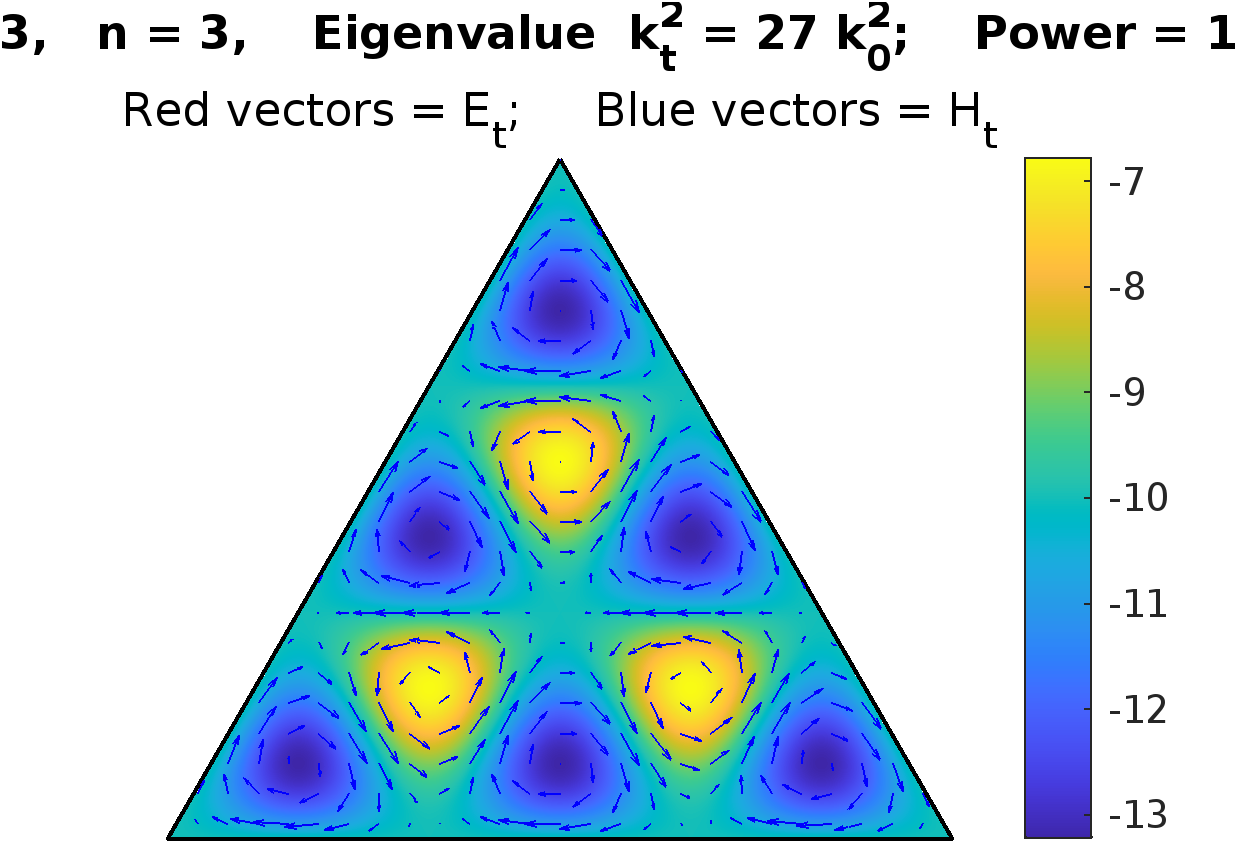}
		\caption{$m=n=3$}
		\label{fig:subfig1}
       \end{minipage}
       \hfill
       \begin{minipage}{0.49\textwidth}
		\includegraphics[width=0.5\linewidth]{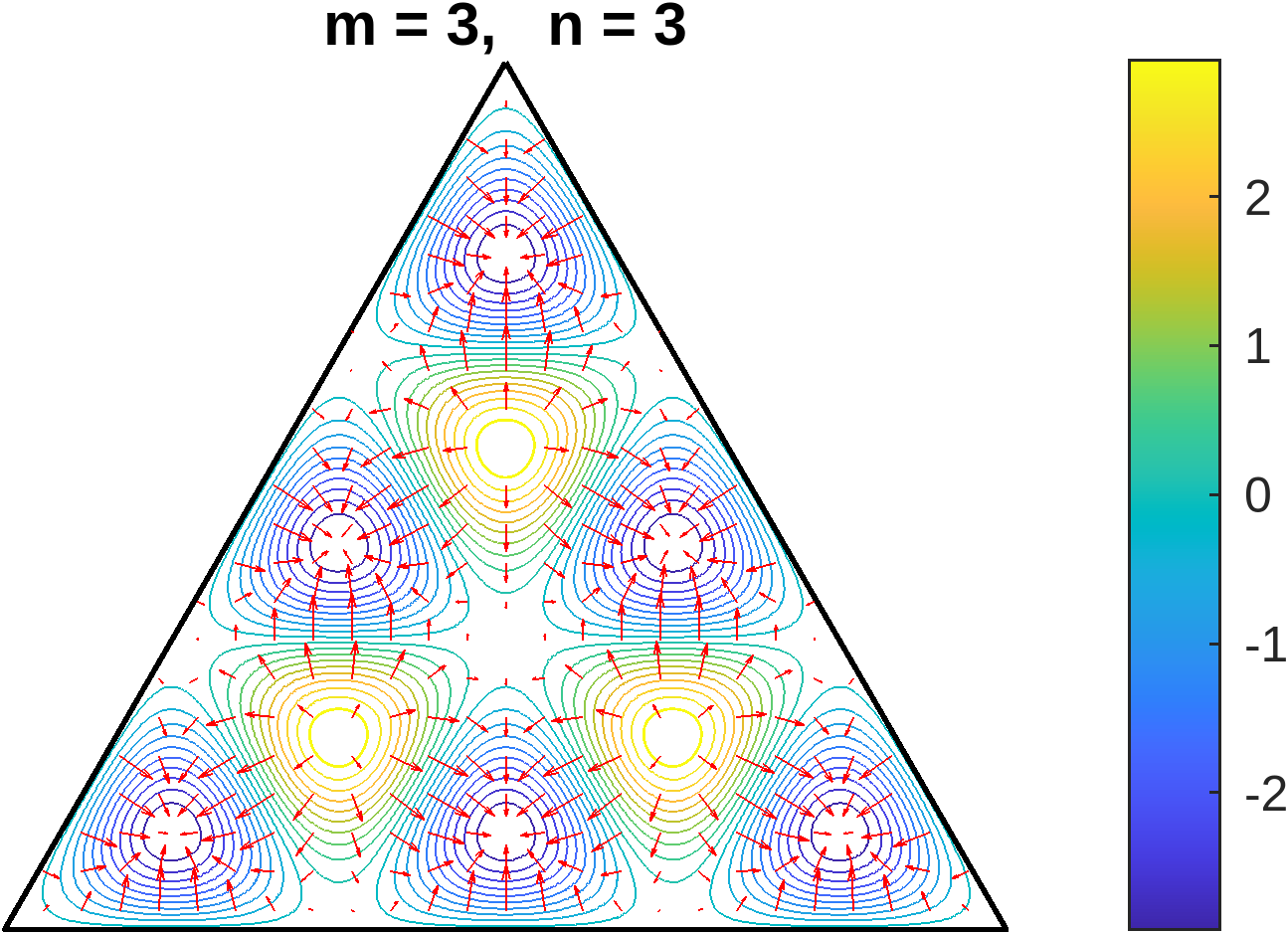}
		\caption{$m=n=3$}
		\label{fig:subfig2}
        \end{minipage}
        
      \begin{minipage}{0.49\textwidth}
		 \includegraphics[width=0.4\linewidth]{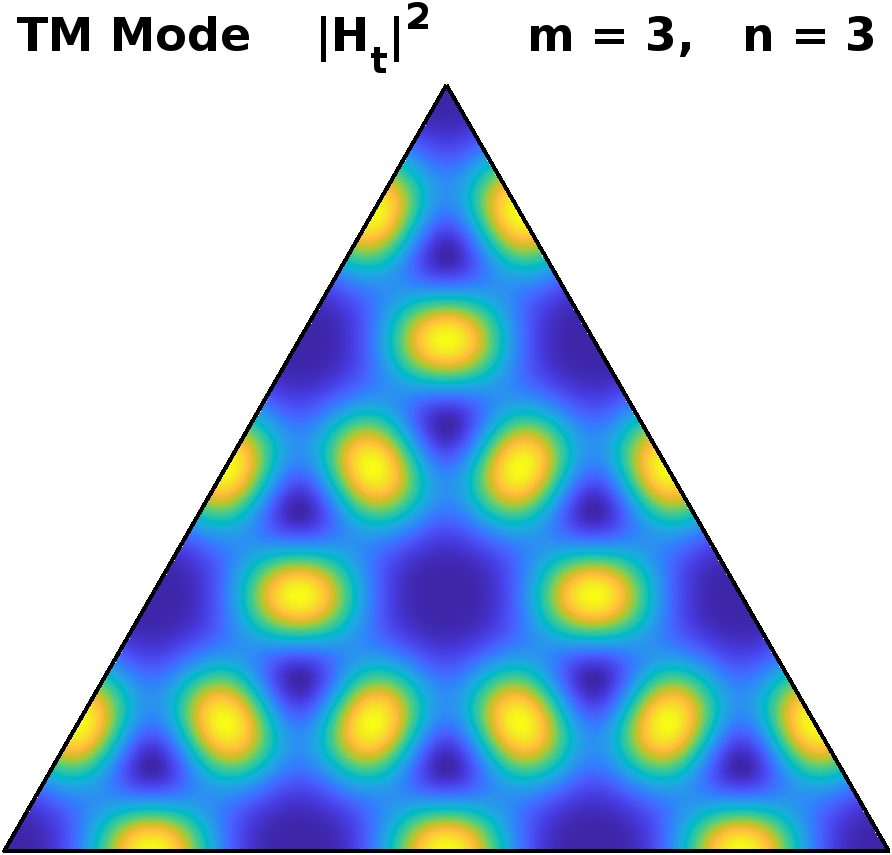}
		 \caption{$m=n=3$}
		 \label{fig:subfig3}
          \end{minipage}
	   \hfil
       \begin{minipage}{0.49\textwidth}
		 \includegraphics[width=0.55\linewidth]{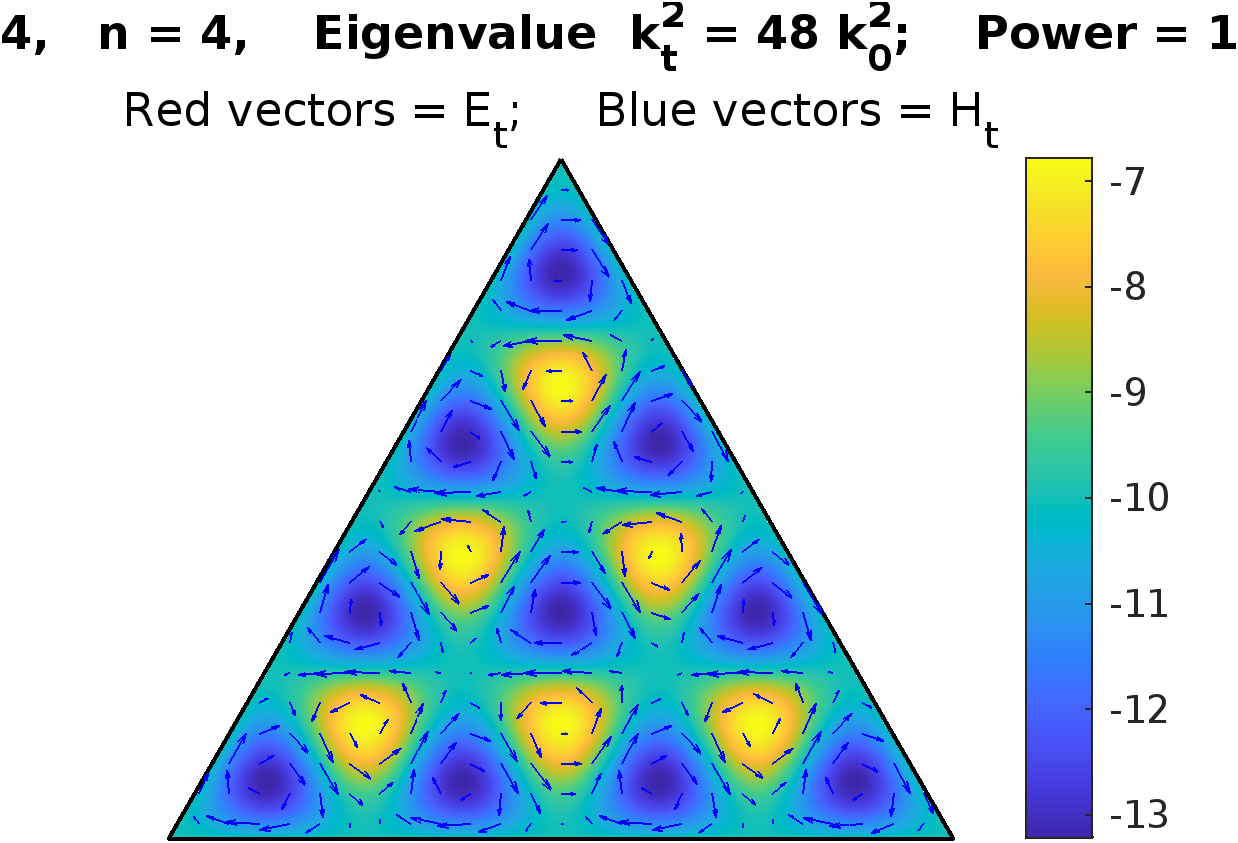}
		 \caption{$m=n=4$}
		 \label{fig:subfig4}
          \end{minipage}
          
      \begin{minipage}{0.49\textwidth}
		  \includegraphics[width=0.55\linewidth]{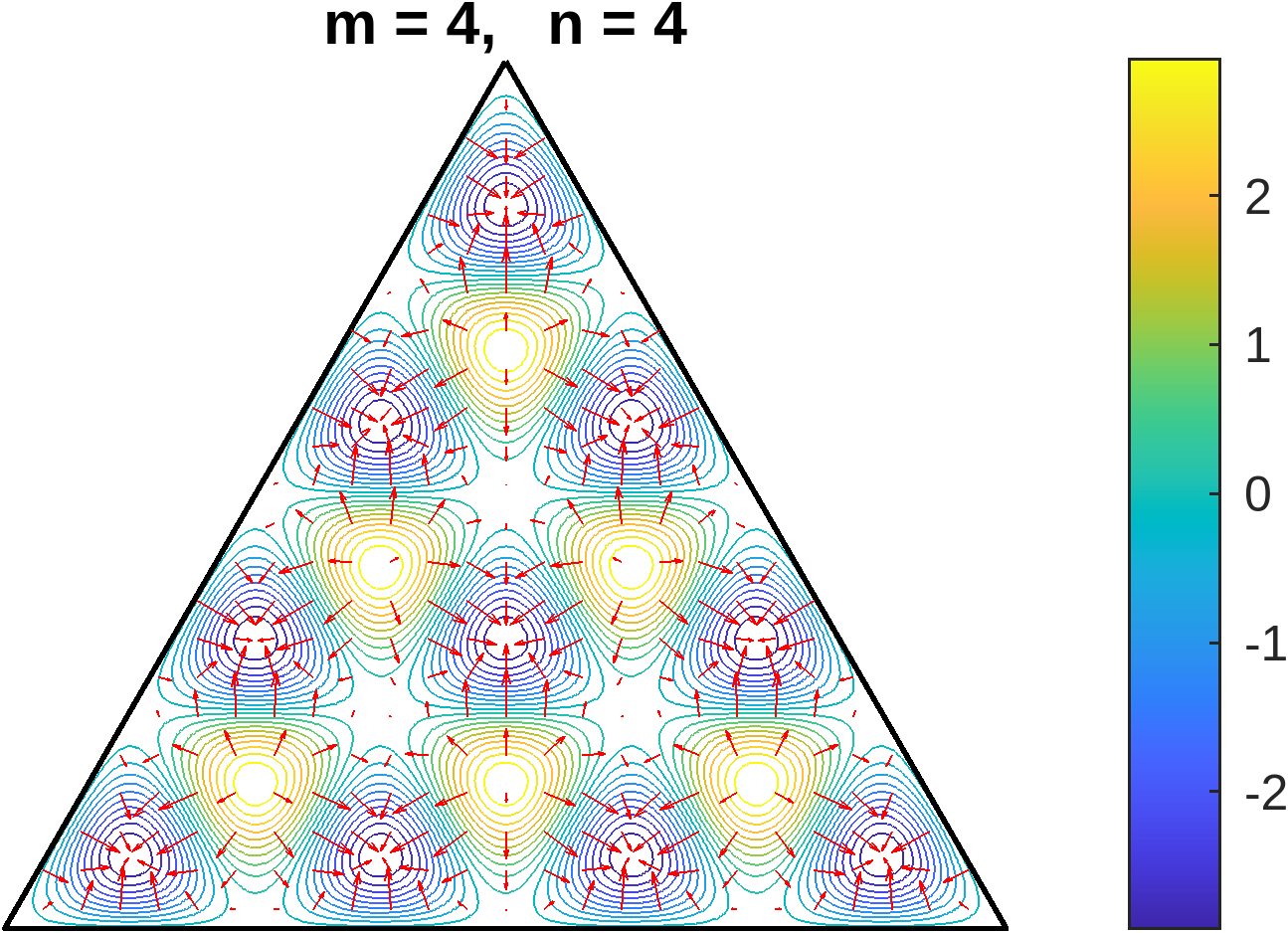}
		  \caption{$m=n=4$}
		  \label{fig:subfig5}
           \end{minipage}
           \hfill
           \begin{minipage}{0.49\textwidth}
		 \includegraphics[width=0.4\linewidth]{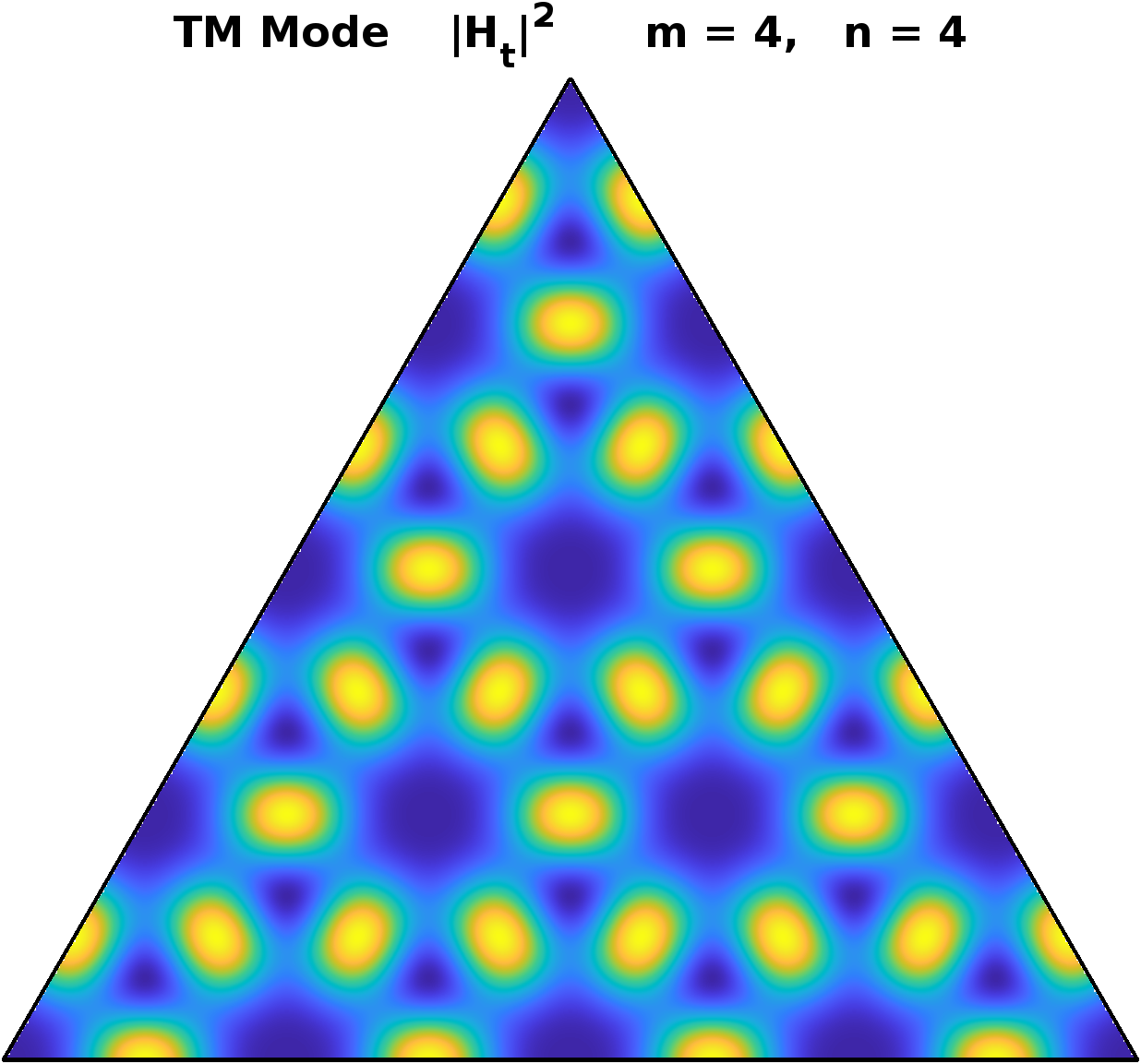}
		 \caption{$m=n=4$}
		 \label{fig:subfig6}
          \end{minipage}
	\caption{Transverse Fields Plots II}
	\label{FigTPII}
\end{figure}

The TM and TE modal structure are shown in fig:\ref{fig:1a}, while their transverse fields are shown in Figs:\ref{FigTP} and \ref{FigTPII}. Our plots compare favorably with the plots found in Alex-Amor et al\cite{AAmor}, even though their plots was wave propagation in equilateral triangular holes, not waveguides. $_{\xi}TM_{lmn}$ and $_{\xi}TE_{lmn}$ denote the TM-modes $_{\xi}\Phi_{lmn}\left(x,y\right)$ and TE-modes $_{\xi}\Psi_{lmn}\left(x,y\right)$ respectively, with $\xi$ denoting the symmetry or parity, which is odd $o$ or even $e$ with respect to the x-coordinate as defined in Eq.s(\ref{E.modes}) and  Eq.s(\ref{H.modes}).

We note that while it is permissible for any two of the indices $l$, $m$ and $n$ to be equal for the TM fields, none of them can be zero, if we want to avoid trivial solutions (we show this in a moment). Thus for the TM fields the lowest possible modes, with the longest cut-off wavelengths (since it has the lowest eigenvalue $k_{\bot}$, to be substituted in $\lambda_{c}=\frac{2\pi}{k_{\bot}}$, or lowest cut-off frequency, since $\omega_{c}=ck_{\bot}=\left(\mu\epsilon\right)^{-\frac{1}{2}}k_{\bot}$) corresponds to $l=-2$, $m=n=1$, or $l=n=1$, $m=-2$, or $l=m=1$, $n=-2$. This particular mode is non-degenerate (with respect to the even or odd solution or the sine-cosine degeneracy\cite{Milton,Borgnis,Onah-PhDthesis}) as we shall see shortly, because two of its indices happen to be equal in this case. The TE fields do not have this restriction, and so as expected in general (for all TE modes), they have the \textit{dominant mode}, with their lowest order corresponding to
$l=0$, $m=1$, $n=-1$, or $l=0$, $m=-1$, $n=1$, or $l=-1$, $m=0$, $n=1$, or $l=-1$, $m=1$, $n=0$, or $l=1$, $m=-1$, $n=0$, or $l=1$, $m=0$, $n=-1$. Where in general
\begin{align}
l+m+n=0.
\label{three.index.relation}
\end{align}

\begin{figure}
\centering
\includegraphics[width=1\textwidth]{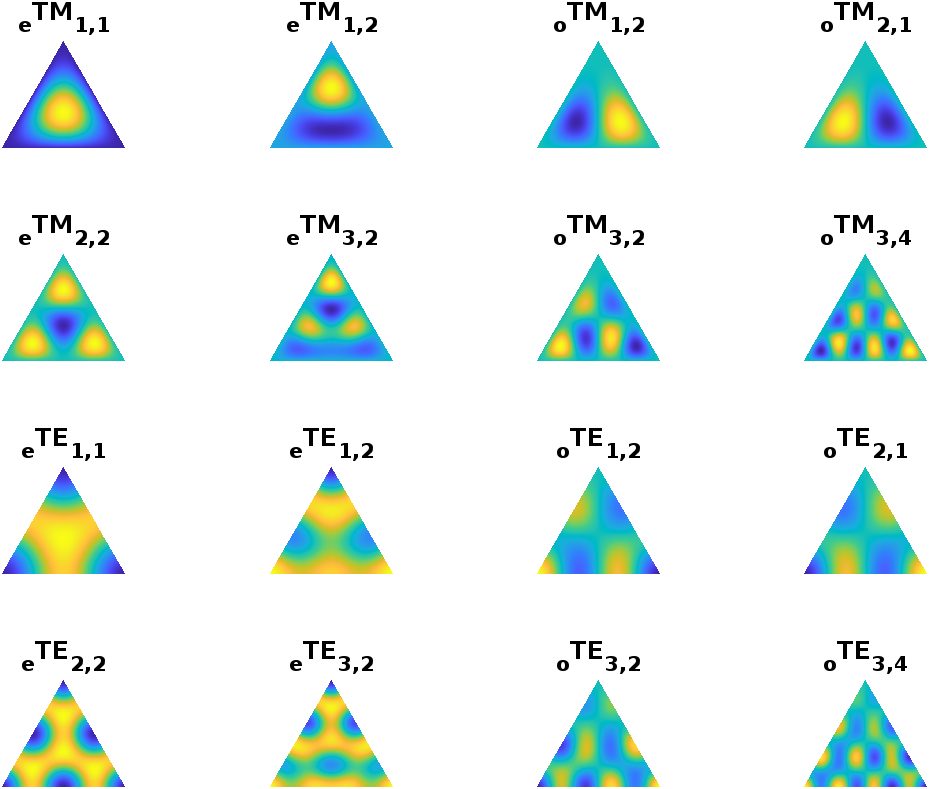}
\caption{figure}{TM and TE-Mode Structure}
\label{fig:1b}
\end{figure}

For equal $m$ and $n$ indices ($l=-m-n=-2n$), the odd modes do not propagate and their intensity plots (of the modes) in such cases are essentially those of the even modes. For any \textit{cyclic permutation} of the indices $l$, $m$ and $n$ (for instance; $_{o}TM_{l,1,2}$, $_{o}TM_{l,2,-3}$ and $_{o}TM_{l,-3,1}$, with $l=-3$, $l=1$ and $l=2$ respectively), we would get the same plots for the odd modes, as also for the even modes (similarly for $_{\xi}TE_{l,m,n}$). See also McCartin's section 3.4\cite{McCartinIII}. 
However, looking at the modes $_{o}TM_{l,1,2}$ and $_{o}TM_{l,2,1}$, we see that \textit{direct} (\textit{odd-permutation}) interchange of $m$ and $n$, leaving $l$ constant results in the same plots for the even modes, but a reflection of the lobes for the odd modes. 
That the interchange of two indices leaving one or the third unchanged, is a symmetry transformation, is a principle we prove shortly. It is interesting to see the flip or interchange of lobes in the modal structures of $_{o}TM_{1,2}$ and $_{o}TM_{2,1}$, and that of $_{o}TE_{1,2}$ and $_{o}TE_{2,1}$, when their indices are interchanged, as shown in fig:\ref{fig:1b}.

In fig:\ref{fig:1a}, we have avoided instances for which $m=n$ in the odd modes (both for the TM and TE modes), since these do not propagate \cite{Milton,Borgnis,McCartinIII} and a surface plot of these modes is simply blank. 

This can readily be seen also from Eq.(\ref{mode.parameters}), from which, for $m=n$, $\theta_{1}=0$ and $\theta_{2}=-\theta_{3}$. Thus, its easy to see in this case, why Eq.s(\ref{E.modes}) and (\ref{H.modes}) are identically zero, and the even modes become (see for instance sec:3.4, pp:39 of McCartin\cite{McCartinIII})
\begin{align}
&_{e}\Phi_{lmn}\left(x,y\right)=2\left[\sin\left(lQ\left(y\right)\right)+2\sin\left(mQ\left(y\right)\right)\cos\left(\theta_{2}x\right)\right],\quad\mathrm{and}\nonumber\\
&_{e}\Psi_{lmn}\left(x,y\right)=2\left[\cos\left(lQ\left(y\right)\right)+2\cos\left(mQ\left(y\right)\right)\cos\left(\theta_{2}x\right)\right].
\end{align}

This is a very well known result\cite{Milton,Borgnis,McCartinII}. It is also not possible for all three indices to be equal, except when they are all equal to zero, since no matter what, the index "$l$" for instance, will always be given by Eq.\ref{three.index.relation} $l=-m-n=-\left(m+n\right)$, in essence, one of the integers will always be equal to minus the sum of the other two. 

Now, we illustrate the odd-permutation of two indices in our mode functions. See that an interchange of the indices "m" and "n" results in new parameters (which we label or denote with a prime), related to that given in (\ref{mode.parameters}) as follows
\begin{align}
&Q^{\prime}\left(y\right)=Q\left(y\right)\quad{;}\quad\theta_{1}^{\prime}=-\theta_{1},\nonumber\\
&\theta_{2}^{\prime}=-\theta_{3}\quad{;}\quad\theta_{3}^{\prime}=-\theta_{2}.
\end{align}

This results in Eq.s(\ref{E.modes}) and (\ref{H.modes}), for $\xi=o$, being odd, in terms of this interchange
\begin{align}
&_{o}\Phi_{lmn}^{\prime}\left(x,y\right)=_{o}\Phi_{lnm}\left(x,y\right)=-_{o}\Phi_{lmn}\left(x,y\right).\nonumber\\
&_{o}\Psi_{lmn}^{\prime}\left(x,y\right)=_{o}\Psi_{lnm}\left(x,y\right)=-_{o}\Psi_{lmn}\left(x,y\right),
\end{align}

while the even modes $\xi=e$, remaining even, with respect to the interchange!
\begin{align}
&_{e}\Phi_{lmn}^{\prime}\left(x,y\right)=_{e}\Phi_{lnm}\left(x,y\right)=_{e}\Phi_{lmn}\left(x,y\right).\nonumber\\
&_{e}\Psi_{lmn}^{\prime}\left(x,y\right)=_{e}\Psi_{lnm}\left(x,y\right)=_{e}\Psi_{lmn}\left(x,y\right).
\end{align}

This is a well known symmetry property\cite{Milton,McCartinIII}, despite the very simple relation Eq.\ref{three.index.relation} $l=-m-n$, between the indices, that may suggest otherwise (since the interchange of $m$ and $n$ gives the same $l$). In essence, just as the change of the sign of "x" in the modal fields Eqs.(\ref{E.modes}) and (\ref{H.modes}) is a \textit{symmetry transformation}, so is the interchange of the indices "m" and "n" leaving "l" unchanged (or any other two combinations, that is $m\leftrightarrow l$ with "n" unchanged or $n\leftrightarrow l$ with "m" unchanged).

Also, for any one of the indices equal to zero (l, say)
\begin{align}
	l=0\quad{;}\quad l+m+n=0\Longrightarrow n=-m,
\end{align}

then from Eq.(\ref{mode.parameters}), we see that
\begin{align}
	&\theta_{2}=\theta_{3}=-\frac{m\pi}{\sqrt{3}h}=\theta,\nonumber\\
	&\theta_{1}=\frac{2m\pi}{\sqrt{3}h}=-2\theta.
\end{align}

So that with
\begin{align}
&\theta_{1}=-2\theta\quad{;}\quad\theta_{2}=\theta_{3}=\theta,\quad\mathrm{and}\nonumber\\
&l=0\quad{;}\quad m=-n,
\end{align}

we see that
\begin{align}	_{e}\Phi_{lmn}\left(x,y\right)=0=_{o}\Phi_{lmn}\left(x,y\right), 
\end{align}

While
\begin{align}	
&_{e}\Psi_{lmn}\left(x,y\right)
=2\left[\cos\left(2\theta x\right)+2\cos\left(mQ\left(y\right)\right)\cos\left(\theta x\right)\right],\quad\mathrm{and}\nonumber\\
&_{o}\Psi_{lmn}\left(x,y\right)=2\left[-\sin\left(2\theta x\right)+2\cos\left(mQ\left(y\right)\right)\sin\left(\theta x\right)\right].
\end{align}

Thus, we have shown that when any of the indices (l, m or n) is zero, the TM modes are always zero or do not propagate. The cyclic and odd permutation properties of the indices illustrated in this section suggest that, while it may be convenient (or elegant) to work with just two indices, courtesy of relation; Eq.\ref{three.index.relation}. It is paramount to retain the third index in some analysis\cite{Milton,Borgnis,McCartinIII}, in fact keeping the third index makes it also apparent why the attenuation constants or quality factors plots for the indices ($m,n$), is precisely the same for \textit{the exact same set} of the indices \textit{with minus signs}: ($-m,-n$), since $l+m+n=0$ and $-l-m-n=0$ are equally valid, but as far as the eigenvalues are concerned, the indices $m$, $n$ and $l$ are all integers. Hence even the negative values are equally valid eigenvalue indices and must be included in the spectrum of the ETW eigenvalues\cite{Milton,Borgnis}, especially as this also has implications for the modal degeneracies(see the last column of table:\ref{eigenvalues.table})\cite{McCartinIII}, which we have seen to be related mostly to the symmetry properties of the mode functions and not just the simple relation in Eq.\ref{three.index.relation}. Besides (\textit{the negative integers}) not being excluded by any physical or mathematical laws, the complication related to the twentieth mode discussed in the next subsection, sometimes the \textit{accidental degeneracies} as we discuss next could also involve negative indices. Also, the eigenvalues which are related to the square root of the factor given in Eq.(\ref{Ensenstein.Form}), have a important significance for either index $m$ or $n$ negative, and the overall value of this factor is never negative, so as to produce imaginary eigenvalues, since this factor was obtained from $l^{2}+m^{2}+n^{2}$, and that is definitely non-negative.    

\subsection{The Eigenvalues}

We have given a detailed ordering of the eigenvalues to the the twentieth mode in table:\ref{eigenvalues.table}, which have been obtained by the analysis of the Eisenstein integers, in particular the Eisenstein form (Eq.\ref{Ensenstein.Form})\cite{McCartinII,Onah-PhDthesis}). The various modes being written in terms of just two indices, is a combination of two such indices out of the three indices into the forms $(m,n)$, $(l,m)$ and $(l,n)$. Obviously, when any two indices are equal, then the index degeneracy is reduced to 2, some $(m,n)$ and $(l,m)$ (say), if $m=n$ (as $(l,m)=(l,n)$).

The need for the number theory of the Eisenstein primes and the \textit{integer prime factorization theorem\cite{GHHardy,DACox,KHRosen}},  becomes important from the 20th mode ($m=3$ and $n=5$, $m=5$ and $n=-8$ and/or $m=-8$ and $n=3$), from which the \textit{accidental degeneracy} becomes apparent. Consider the case for which the indices m and n, say are equal\cite{McCartinII}  $m=7=n$, which implies and eigenvalue of $k_{\bot}=\sqrt{147k_{0}}$. But this gives precisely the same eigenvalue and cut-off frequency, as the mode with indices $m=2$ and $n=11$ or in fact $m=-2$ and $n=13$! Since they have the same Eisenstein form 
 value(and hence eigenvalue)(Eq.\ref{Ensenstein.Form})\cite{McCartinII,McCartinIII}
\begin{align}
E_{m,n}=147&=7^{2}+7^{2}+7^{2}=2^{2}+2\cdot11+11^{2}\nonumber\\
&=\left(-2\right)^{2}-2\cdot13+13^{2}.
\end{align}

Notably, our index combination, as illustrated above does not and cannot reach or connect the two sets of indices with the same eigenvalues (\textit{But as we illustrate shortly, the manipulations of the Eisenstein factor (Eq.\ref{Ensenstein.Form}) and the integer prime factorization theorem does}). It is interesting to note that, while the odd modes do \textit{not} propagate for the mode set $m=7=n$, they do, for the mode sets; $m=2$, $n=11$ and $m=-2$, $n=13$.

Another example is given by $m=5$, $n=6$ and $m=1$, $n=9$,  both of which have $E_{m,n}=91$.

As we have noted, from the 20th mode, given the mode set $m=3$ and $n=5$, it may not come readily to mind that it has the same eigenvalue as the mode sets $m=7$, $n=-7$ and $m=0$, $n=7$.
\begin{align}
E_{m,n}=3^{2}+3\cdot5+5^{2}=49=0^{2}+0\cdot7+7^{2}.
\end{align}

Now, this latest example also has its own complications! For, while the TM modes completely vanish, whenever any of the indices is zero, or when one index is equal to the negative of another ($m=7=-n$ and/or $l=0$), it is not that obvious and in fact, it is not true that the TM modes vanish for $m=3$ and $n=5$ ($l=-8$), which has the same eigenvalue. As the mode ($m=7=-n$ and $l=0$) is present only for the TE mode. However, the TM modes, with this same eigenvalue are present in a different set of indices $m=3$ and $n=5$, which the TE mode also exhibit\cite{Onah-PhDthesis}. Another example of this sort of degeneracy, is that of the pairs $m=7$, $n=8$ (present in the TE and TM modes) and $m=0$, $n=13$ (in which only the TE modes are present), which both have $E_{m,n}=169$.\\ 
The integer prime factorization theorem or \textit{the fundamental theorem of arithmetic} therefore, serves the purpose of catching some of these \textit{not so obvious} or accidental degeneracies\cite{McCartinIII,Onah-PhDthesis}.

Now, we note that as it is sometimes used in the literature, our z-component wave number $\beta$ is $\beta=k_{z}$, while our eigenvalue $k_{\bot}$ is $k_{\bot}=\gamma_{lmn}=k_{t}=\gamma_{mn}$. So that $\gamma_{lmn}^{2}=\omega^{2}\mu\epsilon-k^{2}$ as it is in the literature, and the wave number squared $k^{2}=\omega^{2}\epsilon\mu$, therefore implies that $k^{2}=k_{z}^{2}+\gamma_{mn}^{2}=\beta^{2}+k_{\bot}^{2}$.

\begin{equation}
\fbox{$k_{\bot}^{2}=%
\displaystyle
\frac{16\pi^{2}}{9a^{2}}\left(  m^{2}+mn+n^{2}\right)  $}%
\end{equation}
then%
\begin{equation}
\gamma_{mn}=k_{\bot}=k_{0}\sqrt{m^{2}+mn+n^{2}},\quad k_{0}=\frac{4\pi}{3a}.
\end{equation}

Modes with the same pair $\left(m,n\right)=\left(n,m\right)$ are degenerate. 
\subsection{The Eisenstein Factor}
In view of obtaining the eigenvalues, we write the following product or factor $\frac{1}{2}\left(l^{2}+m^{2}+n^{2}\right)=m^{2}+m\cdot n+n^{2}$ in the Eisenstein form
\begin{align}
E_{m,n}&=\left(m+n\Omega\right)\left(m+n\Bar{\Omega}\right)\nonumber\\
&=m^{2}+m\cdot n+n^{2}=E_{n,m},
\label{Ensenstein.Form}
\end{align}
where
\begin{align}
\Omega=\frac{1}{2}+i\frac{\sqrt{3}}{2},
\end{align}
which is the cube root of $-1$, with the following properties
\begin{align}
&\Omega\Bar{\Omega}=1\quad{;}\quad\Omega^{2}=\Omega-1,\nonumber\\
&\Omega^{2}=\Omega-1\quad{;}\quad\Omega+\Bar{\Omega}=1,\nonumber\\
&\Bar{\Omega}=1-\Omega\quad{;}\quad\Omega^{2}=-\Bar{\Omega},\quad\mathrm{and}\nonumber\\
&\Bar{\Omega}^{2}=-\Omega=\Bar{\Omega}-1.
\label{Eisensetein.root.properties}
\end{align}
One can readily check that $E_{m,n}$ is precisely the same as $E_{m,n}=m^{2}+m\cdot n+n^{2}=\vert m+n\Omega\vert^{2}$(see \cite{McCartinIII}, where $\omega$ is used for $\Omega$ instead). Every such number that can be written in the form of Eq.\ref{Ensenstein.Form} is said to be \textit{Representable} in the literature\cite{McCartinII,McCartinIII}. (Strictly speaking the \textit{Eisenstein Integers} are conventionally defined to be complex numbers that can be represented as $z=m+n\Tilde{\omega}$, where $\Tilde{\omega}=\frac{-1+i\sqrt{3}}{2}$ is the cube root of 1. A modified version of this definition is what we have adopted for our purpose, because a modulus squared of this complex number is actually $\vert z\vert^{2}=\vert m+n\Tilde{\omega}\vert^{2}=m^{2}-m\cdot n+n^{2}$ and not the desired expression that we have in Eq.\ref{Ensenstein.Form}\cite{GHHardy,DACox,KHRosen}). Some examples of representable integers include
\begin{align}
&1=\left(1-\Omega\right)\left(1-\Bar{\Omega}\right),\nonumber\\
&3=\left(1+\Omega\right)\left(1+\Bar{\Omega}\right),\nonumber\\
&7=\left(2+\Omega\right)\left(2+\Bar{\Omega}\right)=\left(1+2\Omega\right)\left(1+2\Bar{\Omega}\right),\nonumber\\
&13=\left(3+\Omega\right)\left(3+\Bar{\Omega}\right)=\left(1+3\Omega\right)\left(1+3\Bar{\Omega}\right),\nonumber\\
&21=\left(4+\Omega\right)\left(4+\Bar{\Omega}\right)=\left(1+4\Omega\right)\left(1+4\Bar{\Omega}\right),\nonumber\\
&31=\left(5+\Omega\right)\left(5+\Bar{\Omega}\right)=\left(1+5\Omega\right)\left(1+5\Bar{\Omega}\right),\nonumber\\
&43=\left(6+\Omega\right)\left(6+\Bar{\Omega}\right)=\left(1+6\Omega\right)\left(1+6\Bar{\Omega}\right),
\end{align}
etc. See that we can obtain infinitely many such representable numbers as $E_{jj}=\left(j+\Omega\right)\left(j+\Bar{\Omega}\right)$ for $j=1,2,3,\cdots$. It is interesting that even the number $1$, has an Eisenstein representation. Note also, that in general $E_{m,n}=E_{n,m}$, as expected. Using Eq.\ref{three.index.relation}, we get other equivalent representations given by $E_{jj}=\left(j-\Omega\right)\left(j-\Bar{\Omega}\right)$ for $j=1,2,3,\cdots$ by the permutation of the three indices $l$, $m$ and $n$.
\begin{align}
&1=\left(0-\Omega\right)\left(0-\Bar{\Omega}\right)=\Omega\Bar{\Omega},\nonumber\\
&3=\left(2-\Omega\right)\left(2-\Bar{\Omega}\right)=\left(1-2\Omega\right)\left(1-2\Bar{\Omega}\right),\nonumber\\
&7=\left(3-\Omega\right)\left(3-\Bar{\Omega}\right)=\left(1-3\Omega\right)\left(1-3\Bar{\Omega}\right),\nonumber\\
&13=\left(4-\Omega\right)\left(4-\Bar{\Omega}\right)=\left(1-4\Omega\right)\left(1-4\Bar{\Omega}\right),\nonumber\\
&21=\left(5-\Omega\right)\left(5-\Bar{\Omega}\right)=\left(1-5\Omega\right)\left(1-5\Bar{\Omega}\right),\nonumber\\
&31=\left(6-\Omega\right)\left(6-\Bar{\Omega}\right)=\left(1-6\Omega\right)\left(1-6\Bar{\Omega}\right),\nonumber\\
&43=\left(7-\Omega\right)\left(7-\Bar{\Omega}\right)=\left(1-7\Omega\right)\left(1-7\Bar{\Omega}\right),
\end{align}

etc. As expected, Eq.\ref{three.index.relation} (\textit{which is a relation originating purely from the equilateral triangular waveguide geometry}) influences or dictates the possible Eisenstein representations or forms for the ETW eigenvalues. Every perfect square have the following general representation (a case for which any of the indices $l$, $m$ or $n$ is zero)
\begin{align}
n^{2}&=\left(n-n\Omega\right)\left(n-n\Bar{\Omega}\right)=\left(n\pm0\cdot\Omega\right)\left(n\pm0\Bar{\Omega}\right)\nonumber\\
&=\left(0\pm n\cdot\Omega\right)\left(0\pm n\cdot\Bar{\Omega}\right),
\end{align}
and in case any two of $l$, $m$ or $n$ are equal ($m=n$, say) we have
\begin{align}
3n^{2}&=\left(n+n\Omega\right)\left(n+n\Bar{\Omega}\right)\nonumber\\
&=\left(n-2n\Omega\right)\left(n-2n\Bar{\Omega}\right)\nonumber\\
&=\left(2n-n\Omega\right)\left(2n-n\Bar{\Omega}\right).
\end{align}
Some other representable forms can also be obtained by combining the smaller representable numbers. For instance, using Eq.\ref{Eisensetein.root.properties}
\begin{align}
21=&3\cdot7=\left(1+\Omega\right)\left(1+\Bar{\Omega}\right)\cdot\left(2+\Omega\right)\left(2+\Bar{\Omega}\right)\nonumber\\
&=\left(1+\Omega\right)\left(2+\Omega\right)\cdot\left(1+\Bar{\Omega}\right)\left(2+\Bar{\Omega}\right)\nonumber\\
&=\left(1+4\Omega\right)\left(1+4\Bar{\Omega}\right),
\end{align}
and similarly for $441=21^{2}=3^{2}\cdot7^{2}$, $39=3\cdot13$, etc. Other representations such as $19=\left(2+3\Omega\right)\left(2+3\Bar{\Omega}\right)$, $37=\left(3+4\Omega\right)\left(3+4\Bar{\Omega}\right)$, etc are also possible, while such numbers as $2$, $5$, $11$, $17$, $23$,... are \textit{Not Representable}, see McCartin\cite{McCartinII,McCartinIII}. \\
Using Eq.\ref{Ensenstein.Form} we see that $k_{\bot}=\sqrt{k_{0}E_{mn}}$.\\

Thus, the frequency at which a mode is cut-off is given by the condition $k^{2}=\gamma_{mn}^{2}=k_{\bot}^{2}$ ($k_{z}=\beta=0$) or
\begin{equation}
0<k_{t}\leq k=\frac{\omega}{c},\qquad\qquad c=\left(\mu_{0}\epsilon
_{0}\right)^{-1/2}=\text{speed of light}%
\end{equation}
then%
\begin{align}
&\gamma_{mn}\leq k=\frac{\omega}{c},\nonumber\\
&\frac{4\pi}{3a}\sqrt{m^{2}+mn+n^{2}}\leq k=\frac{\omega}{c}.
\end{align}

Thus, for the mode $\left(m,n\right)$ to propagate in the guide, its
frequency $\omega$ must be equal to or greater than its cut-off frequency
$\omega_{c}^{\left(m,n\right)}$.%
\begin{align}
\omega\geq\omega_{c}^{\left(m,n\right)}&=c\gamma_{mn}=\frac{4\pi c}{3a}\sqrt{m^{2}+mn+n^{2}}\nonumber\\
&=\omega_{0}\sqrt{E_{m,n}},\nonumber\\
&\quad\mathrm{with}\quad\omega_{0}\equiv
\frac{4\pi c}{3a},
\end{align}
where $\omega_{0}$ is the minimum frequency for having the lowest fundamental
modes: $_{\xi}TE_{0,1}$, $_{\xi}TE_{0,-1}$ and/or $_{\xi}TE_{-1,1}$\cite{Milton,Borgnis,McCartinII}. 

Normalized frequency (with respect to $\omega_{0})$%
\begin{equation}
\tau=\frac{\omega}{\omega_{0}}.
\end{equation}

We close this subsection by showing that when at least one of the mode indices (integers) in one set is representable, we can by manipulating by manipulating the Eisenstein factor (Eq.\ref{Ensenstein.Form}) obtain the other set (or sets), if at least one of those is also representable. For instance, the mode (set) $m=0$ and $n=7$ (where $7$ is representable and using Eq.\ref{Eisensetein.root.properties})
\begin{align}
49&=0^{2}+0\cdot7+7^{2}\nonumber\\
&=7^{2}=\left[\left(2+\Omega\right)\left(2+\Bar{\Omega}\right)\right]^{2}\nonumber\\
&=\left(2+\Omega\right)^{2}\cdot\left(2+\Bar{\Omega}\right)^{2}\nonumber\\
&=\left(3+5\Omega\right)\left(3+5\Bar{\Omega}\right)\nonumber\\
&=3^{2}+3\cdot5+5^{2}.
\end{align}
 So here we get the other mode set: $m=3$ and $n=5$. Finally, taking the case of $m=7=n$ (where $3$ and $7$ are representable)
\begin{align}
147&=7^{2}+7\cdot7+7^{2}\nonumber\\
&=3\cdot7^{2}=\left(1+\Omega\right)\left(1+\Bar{\Omega}\right)\cdot\left[\left(2+\Omega\right)\left(2+\Bar{\Omega}\right)\right]^{2}\nonumber\\
&=\left(1+\Omega\right)\left(2+\Omega\right)^{2}\cdot\left(1+\Bar{\Omega}\right)\left(2+\Bar{\Omega}\right)^{2}\nonumber\\
&=\left(-2+13\Omega\right)\cdot\left(-2+13\Bar{\Omega}\right),
\end{align}

 so that in this case we have obtained the mode set: $m=-2$ and $n=13$. See that no manipulations of the Eisenstein form in this case can lead us to the mode set: $m=2$ and $n=11$, since both $2$ and $11$ are not representable. However, if we recall our old friend $l=-\left(m+n\right)$, we get a third index from $l=-\left(-2+13\right)=-11$, so that another mode set could be $m=-2$ and $n=-11$ (or the familiar $m=2$ and $n=11$) or $m=13$ and $n=-11$: $13^{2}-13\cdot11+\left(-11\right)^{2}=147$.\\
 In these simple examples we have not invoked the
 prime factorization theorem, since the prime factors of the values considered are simple $49=7^{2}$ and $147=3\cdot7^{2}$. Things become more technical, the larger the eigenvalues, especially if some or all of the prime factors are not representable\cite{McCartinII,McCartinIII}.

\begin{center}
\begin{longtable}{|c|c|c|c|c|c|c|}
\caption{in units of $c\sqrt{k_{0}}$, with $\gamma_{mn}=\sqrt{k_{0}}\sqrt{E_{mn}}$ ; $k_{0}=\frac{\pi^{2}}{3h^{2}}$ ;$h=\frac{a\sqrt{3}}{2}$ and $\omega_{c}=ck_{\bot}$.\\ The occasional dashes in the TM modes, indicates absent modes (eigenvalues) for the TM.}
\label{eigenvalues.table}\\
 \hline
 \multicolumn{7}{|c|}{Equilateral Triangular Waveguide Eigenvalues}\\
 \hline
 The Modes & \multicolumn{2}{|c|}{Indices} & $E_{mn}=m^{2}+m\cdot n+n^{2}$ & \multicolumn{2}{|c|}{Eigenvalues ($k_{\bot}=\sqrt{k_{0}E_{mn}}$)} & Degeneracy \\ \hline m,n-Integer values & m & n & $E_{m,n}=E_{n,m}$ & TE & TM ($m\neq0\neq n$) & (Even-Odd)\\
 \hline 
 \endfirsthead
 
 \hline
 \multicolumn{7}{|c|}{Table\thetable{}--continued from previous page}\\
 \hline
 \endhead
 
 \hline
 \multicolumn{7}{|r|}{Continued on next page}\\
 \hline
 \endfoot
 
 \hline
 \multicolumn{7}{|r|}{End of Table\thetable}\\
 \hline
 \endlastfoot
 
 \hline
 \multirow{3}{3em}{First Mode} & $1$ & $-1$ & $2$ & $\sqrt{2}$ & $-$ & $TE^{e,o}_{-1,1}$\\
 & $0$ & $1$ & $2$ &  $\sqrt{2}$ & $-$ & $TE_{0,1}^{e,o}$ \\
& $-1$ &$0$ & $2$ &  $\sqrt{2}$ & $-$ & $TE^{e,o}_{-1,0}$\\
 \hline
\multirow{2}{3em}{Second Mode} & $1$ &  $1$ & $3$ & $\sqrt{3}$ & $\sqrt{6}$ & $TE^{e}_{1,1}$,$TM^{e}_{1,1}$\\
 & $-2$ & $1$ & $3$ & $\sqrt{3}$ & $\sqrt{6}$ & $TE^{e}_{-2,1}$,$TM^{e}_{-2,1}$\\
 \hline
 \multirow{3}{3em}{Third Mode} & $0$ & $2$ & $4$ & $2$ & $-$ & $TE^{e,o}_{0,2}$\\
 & $-2$ & $2$ & $4$ & $2$ & $-$ & $TE^{e,o}_{-2,2}$ \\
 & $0$ & $-2$ & $4$ & $2$ & $-$ & $TE^{e,o}_{0,-2}$\\
 \hline
\multirow{3}{3em}{Forth Mode} & $1$ & $2$ & $7$ & $\sqrt{7}$ & $\sqrt{14}$ & $TE^{e,o}_{1,2}$,$TM^{e,o}_{1,2}$\\
 & $-3$ & $1$ & $7$ & $\sqrt{7}$ & $\sqrt{14}$ & $TE^{e,o}_{-3,1}$,$TM^{e,o}_{-3,1}$\\
 & $2$ & $-3$ & $7$ & $\sqrt{7}$ & $\sqrt{14}$ & $TE^{e,o}_{2,-3}$,$TM^{e,o}_{2,-3}$ \\
 \hline
 \multirow{3}{2em}{Fifth Mode} & $3$ & $0$ & $9$ & $3$ & $-$ & $TE^{e,o}_{3,0}$\\
 & $-3$ & $3$ & $9$ & $3$ & $-$ & $TE^{e,o}_{-3,3}$\\
 & $-3$ & $0$ & $9$ & $3$ & $-$ & $TE^{e,o}_{-3,0}$ \\
 \hline
 \multirow{2}{3em}{Sixth Mode} & $2$ & $2$ & $12$ & $2\sqrt{3}$ & $2\sqrt{3}$ & $TE^{e}_{2,2}$,$TM^{e}_{2,2}$\\
 & $-4$ & $2$ & $12$ & $2\sqrt{3}$ & $2\sqrt{3}$ & $TE^{e}_{-4,2}$,$TM^{e}_{-4,2}$\\
 \hline
 \multirow{3}{3em}{Seventh Mode} & $1$ & $3$ & $13$ & $\sqrt{13}$ & $\sqrt{13}$ & $TE^{e,o}_{1,3}$,$TM^{e,o}_{1,3}$\\
 & $-4$ & $3$ & $13$ & $\sqrt{13}$ & $\sqrt{13}$ & $TE^{e,o}_{-4,3}$,$TM^{e,o}_{-4,3}$\\
 & $1$ & $-4$ & $13$ & $\sqrt{13}$ & $\sqrt{13}$ & $TE^{e,o}_{1,-4}$,$TM^{e,o}_{1,-4}$ \\
 \hline
 \multirow{3}{3em}{Eighth Mode} & $0$ & $4$ & $16$ & $4$ & $-$ & $TE^{e,o}_{0,4}$\\
 & $-4$ & $4$ & $16$ & $4$ & $-$ & $TE^{e,o}_{-4,4}$\\
 & $-4$ & $0$ & $16$ & $4$ & $-$ & $TE^{e,o}_{-4,0}$ \\
 \hline
 \multirow{3}{3em}{Ninth Mode} & $3$ & $2$ & $19$ & $\sqrt{19}$ & $\sqrt{19}$ & $TE^{e,o}_{3,2}$,$TM^{e,o}_{3,2}$\\
 & $-5$ & $3$ & $19$ & $\sqrt{19}$ & $\sqrt{19}$ & $TE^{e,o}_{-5,3}$,$TM^{e,o}_{-5,3}$\\
 & $2$ & $-5$ & $19$ & $\sqrt{19}$ & $\sqrt{19}$ & $TE^{e,o}_{2,-5}$.$TM^{e,o}_{2,-5}$ \\
 \hline
 \multirow{3}{3em}{Tenth Mode} & $1$ & $4$ & $21$ & $\sqrt{21}$ & $\sqrt{21}$ & $TE^{e,o}_{1,4}$,$TM^{e,o}_{1,4}$\\
 & $-5$ & $4$ & $21$ & $\sqrt{21}$ & $\sqrt{21}$ & $TE^{e,o}_{-5,4}$,$TM^{e,o}_{-5,4}$\\
 & $1$ & $-5$ & $21$ & $\sqrt{21}$ & $\sqrt{21}$ & $TE^{e,o}_{2,-5}$,$TM^{e,o}_{2,-5}$ \\
 \hline
 \multirow{3}{3em}{Eleventh Mode} & $0$ & $5$ & $25$ & $5$ & $-$ & $TE^{e,o}_{0,5}$\\
 & $-5$ & $5$ & $25$ & $5$ & $-$ & $TE^{e,o}_{-5,5}$\\
 & $0$ & $-5$ & $25$ & $5$ & $-$ & $TE^{e,o}_{0,-5}$ \\
 \hline
 \multirow{2}{2em}{Twelfth Mode} & $3$ & $3$ & $27$ & $\sqrt{27}$ & $\sqrt{27}$ & $TE^{e}_{3,3}$,$TM^{e}_{3,3}$\\
 & $-6$ & $3$ & $27$ & $\sqrt{27}$ & $\sqrt{27}$ & $TE^{e}_{-6,3}$,$TM^{e}_{-6,3}$\\
 \hline
 \multirow{3}{3em}{Thirteenth Mode} & $2$ & $4$ & $28$ & $2\sqrt{6}$ & $2\sqrt{6}$ & $TE^{e,o}_{2,4}$,$TM^{e,o}_{2,4}$\\
 & $-6$ & $4$ & $28$ & $2\sqrt{6}$ & $2\sqrt{6}$ & $TE^{e,o}_{-6,4}$.$TM^{e,o}_{-6,4}$\\
 & $2$ & $-6$ & $28$ & $2\sqrt{6}$ & $2\sqrt{6}$ & $TE^{e,o}_{2,-6}$,$TM^{e,o}_{2,-6}$ \\
 \hline
 \multirow{3}{3em}{Fourteenth Mode} & $1$ & $5$ & $31$ & $\sqrt{31}$ & $\sqrt{31}$ & $TE^{e,o}_{1,5}$,$TM^{e,o}_{1,5}$\\
 & $-6$ & $1$ & $31$ & $\sqrt{31}$ & $\sqrt{31}$ & $TE^{e,o}_{-6,1}$,$TM^{e,o}_{-6,1}$\\
 & $5$ & $-6$ & $31$ & $\sqrt{31}$ & $\sqrt{31}$ & $TE^{e,o}_{5,-6}$,$TM^{e,o}_{5,-6}$ \\
 \hline
 \multirow{3}{3em}{Fifteenth Mode} & $0$ & $6$ & $36$ & $6$ & $-$ & $TE^{e,o}_{0,6}$\\
 & $-6$ & $6$ & $36$ & $6$ & $-$ & $TE^{e,o}_{-6,6}$\\
 & $0$ & $-6$ & $36$ & $6$ & $-$ & $TE^{e,o}_{0,-6}$ \\
 \hline
 \multirow{3}{3em}{Sixteenth Mode} & $3$ & $4$ & $37$ & $\sqrt{37}$ & $\sqrt{37}$ & $TE^{e,o}_{3,4}$,$TM^{e,o}_{3,4}$\\
 & $-7$ & $4$ & $37$ & $\sqrt{37}$ & $\sqrt{37}$ & $TE^{e,o}_{-7,4}$,$TM^{e,o}_{-7,4}$\\
 & $3$ & $-7$ & $37$ & $\sqrt{37}$ & $\sqrt{37}$ & $TE^{e,o}_{3,-7}$,$TM^{e,o}_{3,-7}$ \\
 \hline
 \multirow{3}{3em}{Seventeenth Mode} & $2$ & $5$ & $39$ & $\sqrt{39}$ & $\sqrt{39}$ & $TE^{e,o}_{2,5}$,$TM^{e,o}_{2,5}$\\
 & $-7$ & $2$ & $39$ & $\sqrt{39}$ & $\sqrt{39}$ & $TE^{e,o}_{-7,2}$,$TM^{e,o}_{-7,2}$\\
 & $5$ & $-7$ & $39$ & $\sqrt{39}$ & $\sqrt{39}$ & $TE^{e,o}_{5,-7}$,$TM^{e,o}_{5,-7}$ \\
 \hline
 \multirow{3}{3em}{Eighteenth Mode} & $1$ & $6$ & $43$ & $\sqrt{43}$ & $\sqrt{43}$ & $TE^{e,o}_{1,6}$,$TM^{e,o}_{1,6}$\\
 & $-7$ & $1$ & $43$ & $\sqrt{43}$ & $\sqrt{43}$ & $TE^{e,o}_{-7,1}$,$TM^{e,o}_{-7,1}$\\
 & $6$ & $-7$ & $43$ & $\sqrt{43}$ & $\sqrt{43}$ & $TE^{e,o}_{6,-7}$,$TM^{e,o}_{6,-7}$ \\
 \hline
 \multirow{3}{3em}{Nineteenth Mode} & $4$ & $4$ & $48$ & $\sqrt{48}$ & $\sqrt{48}$ & $TE^{e}_{4,4}$,$TM^{e}_{4,4}$\\
 & $-8$ & $4$ & $48$ & $\sqrt{48}$ & $\sqrt{48}$ & $TE^{e}_{-8,4}$.$TM^{e}_{-8,4}$\\
 \hline
 \multirow{3}{3em}{Twentieth Mode I} & $3$ & $5$ & $49$ & $7$ & $7$ & $TE^{e,o}_{3,5}$,$TM^{e,o}_{3,5}$\\
 & $-8$ & $3$ & $49$ & $7$ & $7$ & $TE^{e,o}_{-8,3}$,$TM^{e,o}_{-8,3}$\\
 & $5$ & $-8$ & $49$ & $7$ & $7$ & $TE^{e,o}_{5,-8}$,$TM^{e,o}_{5,-8}$ \\
 \hline
 \multirow{3}{3em}{Twentieth Mode II} & $0$ & $7$ & $49$ & $7$ & $-$ & $TE^{e,o}_{0,7}$\\
 & $-7$ & $0$ & $49$ & $7$ & $-$ & $TE^{e,o}_{-7,0}$\\
 & $7$ & $-7$ & $49$ & $7$ & $-$ & $TE^{e,o}_{7,-7}$\\
\hline
\end{longtable}
\end{center}

\section{ETW Attenuation Constants}

The standard relations for the E- and H-mode attenuation constants are given by
\begin{align}
&\alpha^{(TM)}=\frac{\omega^{2}\epsilon\mu\delta_{s}}{4kk_{\bot}^{2}}\frac{\oint_{C}\vert\mathbf{\nabla}_{\bot}\Phi_{lmn}\left(x,y\right)\vert^{2}dl}{\int_{A}\vert\Phi_{lmn}\left(x,y\right)\vert^{2}da},\nonumber\\
&\alpha^{(TE)}=\frac{\delta_{s}}{4kk_{\bot}^{2}}\frac{\oint_{C}\left(k_{\bot}^{4}\vert\Psi_{lmn}\left(x,y\right)\vert^{2}+k^{2}\vert\mathbf{\nabla}_{\bot}\Psi_{lmn}\left(x,y\right)\vert^{2}\right)dl}{\int_{A}\vert\Psi_{lmn}\left(x,y\right)\vert^{2}da},
\end{align}

where (see also \cite{YZHuang})
\begin{align}
&\Phi_{lmn}\left(x,y\right)=_{e}\Phi_{lmn}\left(x,y\right)+i\left(_{o}\Phi_{lmn}\left(x,y\right)\right),\nonumber\\
&\Psi_{lmn}\left(x,y\right)=_{e}\Psi_{lmn}\left(x,y\right)+i\left(_{o}\Psi_{lmn}\left(x,y\right)\right),
\label{eigenfunctions.super}
\end{align}

but we have used the normalized longitudinal fields of Borgnis and Papas\cite{Borgnis}, stated as
\begin{align}
&\int_{A}\mathbf{\nabla}_{\bot}\Phi_{p}\left(x,y\right)\cdot\mathbf{\nabla}_{\bot}\Phi_{q}\left(x,y\right)da=\delta_{pq},\nonumber\\ &\int_{A}\mathbf{\nabla}_{\bot}\Psi_{p}\left(x,y\right)\cdot\mathbf{\nabla}_{\bot}\Psi_{q}\left(x,y\right)da=\delta_{pq},
\end{align}

which implies that
\begin{align}
\int_{A}\vert\Phi_{lmn}\left(x,y\right)\vert^{2}da=\frac{1}{k_{\bot}^{2}}\quad\mathrm{and}\quad\int_{A}\vert\Psi_{lmn}\left(x,y\right)\vert^{2}da=\frac{1}{k_{\bot}^{2}}.
\end{align}

Thus with
\begin{align}	I_{i}^{(TM)}=\int_{c_{i}}\vert\mathbf{\nabla}_{\bot}\Phi_{lmn}\left(x,y\right)\vert^{2}dl_{i}\quad{;}\quad i=1,2,3.
\end{align}

And
\begin{align}
&I_{1i}^{(TE)}=\int_{c_{i}}\vert\Psi_{lmn}\left(x,y\right)\vert^{2}dl_{i}\nonumber\\ &I_{2i}^{(TE)}=\int_{c_{i}}\vert\mathbf{\nabla}_{\bot}\Psi_{lmn}\left(x,y\right)\vert^{2}dl_{i}\quad{;}\quad i=1,2,3.
\end{align}

The \textit{normalized attenuation constants} in terms of \textit{normalized frequencies} are given by
\begin{align}
\alpha^{\prime (TM)}=\sqrt{\frac{\tau_{1}^{3}}{\tau_{1}^{2}-1}}\left(\sqrt{k_{\bot}}\oint_{C}\vert\vec{\nabla}_{\bot}\Phi_{lmn}\left(x,y\right)\vert^{2}dl\right),
\label{General.Nor.TM.Att}
\end{align}

and,
\begin{align}
\alpha^{\prime(TE)}&=\frac{1}{\sqrt{\tau_{2}^{3}-\tau_{2}}}\left(k_{\bot}^{\frac{5}{2}}\oint_{C}\vert\Psi_{lmn}\left(x,y\right)\vert^{2}dl\right)\nonumber\\
&+\sqrt{\frac{\tau_{2}^{2}-1}{\tau_{2}}}\left(\sqrt{k_{\bot}}\oint_{C}\vert\vec{\nabla}_{\bot}\Psi_{lmn}\left(x,y\right)\vert^{2}dl\right),
\label{General.Nor.TE.Att}
\end{align}
Eqs(\ref{General.Nor.TM.Att}) and (\ref{General.Nor.TE.Att}) are general. They are the general normalized attenuation constants for any cylindrical waveguide. Thus for the ETW we have, specifically 
\begin{align}
\alpha^{\prime(TM)}=\sqrt{\frac{\tau_{1}^{3}}{\tau_{1}^{2}-1}}\sqrt{k_{\bot}}\left[I_{1}^{(TM)}+I_{2}^{(TM)}+I_{3}^{(TM)}\right],
\label{attenuationIIa}
\end{align}
\begin{align}
\alpha^{\prime(TE)}&=\frac{1}{\sqrt{\tau_{2}^{3}-\tau_{2}}}k_{\bot}^{\frac{5}{2}}\left[I_{11}^{(TE)}+I_{12}^{(TE)}+I_{13}^{(TE)}\right]\nonumber\\
&+\sqrt{\frac{\tau_{2}^{2}-1}{\tau_{2}}}\sqrt{k_{\bot}}\left[I_{21}^{(TE)}+I_{22}^{(TE)}+I_{23}^{(TE)}\right],
\label{attenuationIIb}
\end{align}

where\cite{NoriegaM}
\begin{align}
&\alpha^{\prime(TM)}=\frac{\alpha^{(TM)}}{e}\left[\frac{2\sigma_{2c}\mu_{2c}Z}{\mu}\right]^{\frac{1}{2}}\quad{;}\quad\alpha^{\prime(TE)}=\frac{\alpha^{(TE)}}{e}\left[\frac{2\sigma_{2c}\mu_{2c}Z}{\mu}\right]^{\frac{1}{2}}\nonumber\\
&\mathrm{with}\quad Z=\sqrt{\frac{\mu}{\epsilon}}.\quad\mathrm{The}\quad\mathrm{Scaling}\quad\mathrm{factor}\quad e=10^{7}.
\end{align}

and
\begin{align}
&\tau_{1}=\frac{\omega}{_{\xi}\omega_{c}^{(TM)}}\quad{;}\quad\tau_{2}=\frac{\omega}{_{\xi}\omega_{c}^{(TE)}},\nonumber\\
&_{\xi}\omega_{c}^{(TM)}=\frac{_{\xi}k_{\bot}^{(TM)}}{\sqrt{\mu\epsilon}}\quad{;}\quad_{\xi}\omega_{c}^{(TE)}=\frac{_{\xi}k_{\bot}^{(TE)}}{\sqrt{\mu\epsilon}},
\end{align}

with
\begin{align}
_{\xi}\omega_{c}=c\left(_{\xi}k_{\bot}\right)\quad{;}\quad\left(_{\xi}\lambda_{c}\right)=\frac{2\pi}{_{\xi}k_{\bot}}\quad{;}\quad_{\xi}\omega_{c}\left(_{\xi}\lambda_{c}\right)=2\pi c.
\end{align}

\begin{figure}
\centering
\includegraphics[width=0.7\textwidth]{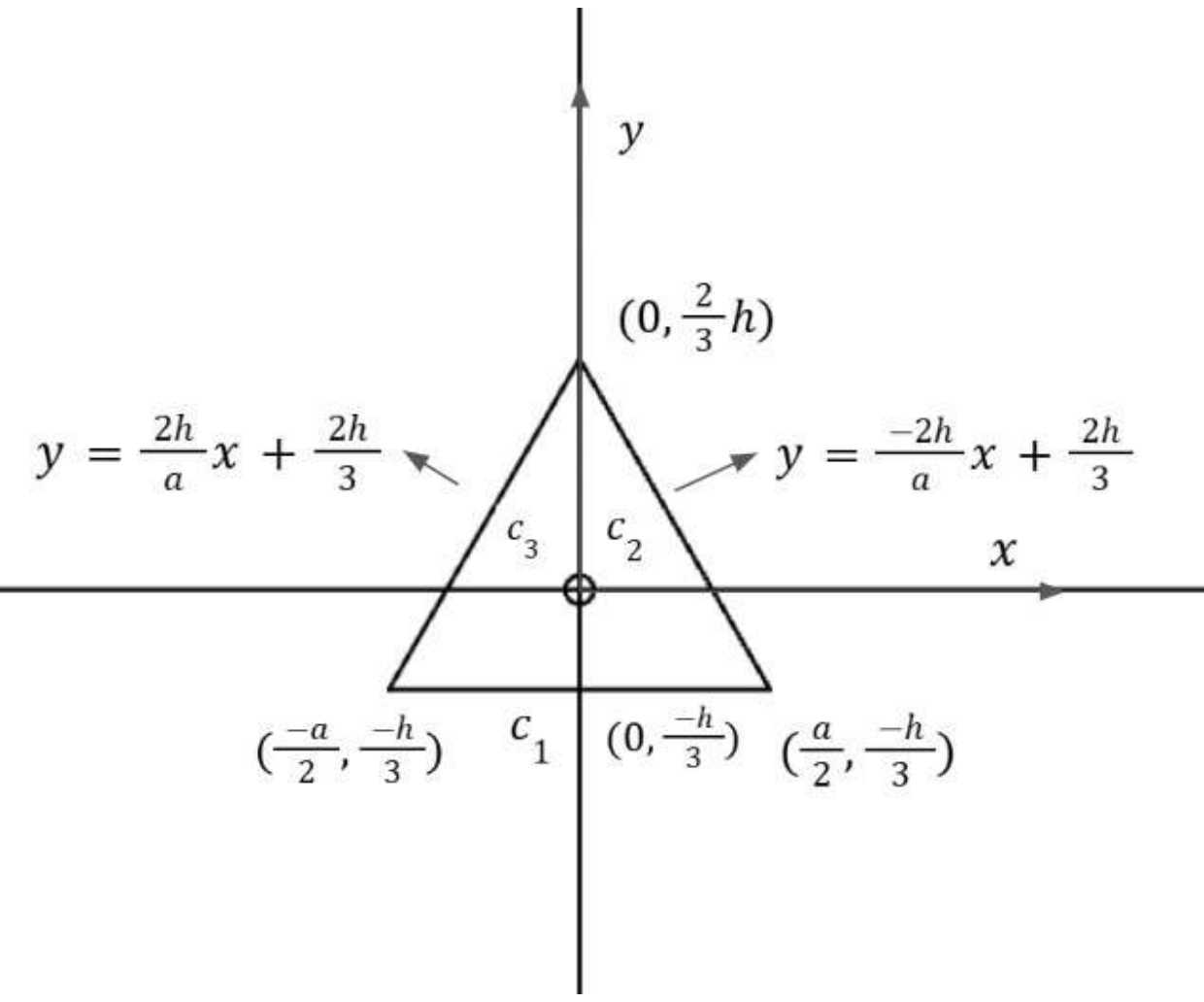}
\caption{An Equilateral Triangular Waveguide Cross Section}
\label{ETWsketchII}
\end{figure}

The specific contours used are for an equilateral triangle of height $h=\frac{\sqrt{3}}{2}a$ and centered around the origin, with the base of the triangle at distance $y=-\frac{h}{3}$ below the origin, just as shown in Fig:\ref{ETWsketchII}. It is important to mention also that the eigen functions of Eq.\eqref{E.modes} and Eq.\eqref{H.modes} were derived from this contour (of Fig:\ref{ETWsketchII}) particularly with the important requirement that the TM modes and normal derivative of the TE modes vanish at the point $u=y=-r=-\frac{h}{3}$ (the origin of the Cartesian coordinates is located at the center of the equilateral triangle with an inscribed circle of radius $r=\frac{h}{3}$, or radius $-r$ in trilinear coordinates), just as captured in Fig\ref{ETWsketch}.

\textit{Now by this it should be clear that the initial error related to the contours $c_{2}$ and $c_{3}$ (and hence also to the parameter $t_{1}$) in the published version of F. E. Onah\cite{Onah2026} are now corrected, so that the integrals  $I^{(TM)}_{2}=I^{(TM)}_{3}$, $I^{(TE)}_{12}=I^{(TE)}_{13}$ and $I^{(TE)}_{22}=I^{(TE)}_{23}$ given in \cite{Onah2026} should read as presented in the appendix of the current version of this document. It is important to state clearly in assuring the reader that all major conclusions, observations (all of which can be confirmed and reproduced) and other derivations in F. Emenike Onah\cite{Onah2026} are correct and accurately represented.} 

The initial error in \cite{Onah2026} consists in the over-estimation of the triangular height (by $\frac{h}{3}$, with $h=a\frac{\sqrt{3}}{2}$) essentially making it an isosceles triangle (of one side (the base) length $a$ and the other two equal sides of length $a\sqrt{\frac{19}{12}}$ and of height $\frac{2a}{\sqrt{3}}$) instead of an equilateral triangle as originally intended. So the integrals as presented in the published article are correct (to best of the author's knowledge), but because the contours used therein \cite{Onah2026} were essentially (in error) those of an isosceles triangular cross section, thus the integrals  $I^{(TM)}_{2}=I^{(TM)}_{3}$, $I^{(TE)}_{12}=I^{(TE)}_{13}$ and $I^{(TE)}_{22}=I^{(TE)}_{23}$ as given therein would differ from the ones you see here in this revised document (which are now simpler). 

\textit{All integrals calculated for the first contour $c_{1}$ in the paper\cite{Onah2026} are correct and are part of our equilateral triangular cross section and contour}. As has been tested and expected of any cylindrical electromagnetic waveguides, the attenuation constant and quality factor plots are the same, except for the overall scaling of the plots given. The notable difference here happen to be the quoted minima attenuation constants in the published version\cite{Onah2026}, which must change because of the change in integrals. Of course the analysis of other plots, modal structures, eigenvalues, Eisenstein primes, vertex symmetries and pizza structures were completely independent of the contours or integrals given, so those stand rock solid as can be verified or reproduced by anyone.

Finally, we update the quoted attenuation minima given in F. E. Onah\cite{Onah2026}. For the mode $m=1=n$, a TM minimum attenuation of $\alpha^{\prime(TM)}_{min}=0.27424$ occurs at $\tau_{1}=\sqrt{3}$, while the corresponding TE minimum attenuation of $\alpha^{\prime(TE)}_{min}=0.340278$ occurs at $\tau_{2}=\sqrt{2}+1=\sqrt{3+2\sqrt{2}}=2.4142135624$. Generally as reported\cite{Onah2026}, all TM minima occur at $\tau_{1}=\sqrt{3}$, while the TE minima attenuation constants depends on the parameter $g$. The TE attenuation minima generally oscillates between the frequency $\tau_{2}=\sqrt{2}+1=\sqrt{3+2\sqrt{2}}$, for non-zero $m$ and $n$ indices, and $\tau_{2}=4.86832$, strictly for at least $m=0$ (and $n\neq0$) or $n=0$ (and $m\neq0$).

\begin{align}
c_{1}:&\quad dl_{1}=dx\quad{;}\quad y=-\frac{h}{3},\nonumber\\
&Q\left(y\right)=-\pi\quad{;}\quad-\frac{a}{2}\leq x\leq\frac{a}{2},
\end{align}

\begin{align}
c_{2}:&\quad dl_{2}=t_{1}dx\quad{;}\quad y=-\frac{2h}{a}x+\frac{2h}{3},\nonumber\\
&Q\left(y\right)=-\frac{2\pi}{a}x\quad{;}\quad \frac{a}{2}\leq x\leq0,
\end{align}

\begin{align}
c_{3}:&\quad dl_{3}=t_{1}dx\quad{;}\quad y=\frac{2h}{a}+\frac{2h}{3},\nonumber\\
 &Q\left(y\right)=\frac{2\pi}{a}x\quad{;}\quad0\leq x\leq-\frac{a}{2}, 
\end{align}

with
\begin{align}
t_{1}=\frac{1}{a}\left(a^{2}+4h^{2}\right)^{\frac{1}{2}}.
\end{align}

Where the superposition given in Eq.\eqref{eigenfunctions.super} implies
\begin{align}
&\Phi_{lmn}\left(x,y\right)=2i\left[\sin\left(lQ\left(y\right)\right)e^{i\theta_{1}x}+\sin\left(mQ\left(y\right)\right)e^{i\theta_{2}x}+\sin\left(nQ\left(y\right)\right)e^{i\theta_{3}x}\right],\nonumber\\
&\Psi_{lmn}\left(x,y\right)=2\left[\cos\left(lQ\left(y\right)\right)e^{i\theta_{1}x}+\cos\left(mQ\left(y\right)\right)e^{i\theta_{2}x}+\cos\left(nQ\left(y\right)\right)e^{i\theta_{3}x}\right],
\end{align}

the modulus square products implies for instance
\begin{align}
\vert\Phi_{lmn}\left(x,y\right)\vert^{2}=\Phi_{lmn}\left(x,y\right)\Phi_{lmn}\left(x,y\right)^{*}\quad{;}\quad\vert\vec{\nabla}_{\bot}\Phi_{lmn}\left(x.y\right)\vert^{2}=\vec{\nabla}_{\bot}\Phi_{lmn}\left(x,y\right)\cdot\vec{\nabla}_{\bot}\Phi_{lmn}\left(x,y\right)^{*}.
\end{align}

We present the explicit expressions of the integrals appearing in Eq.s(\ref{attenuationIIa}),(\ref{attenuationIIb}) and (\ref{TMQuality.factorII}),(\ref{TEQuality.factorII}) for the attenuation constants and quality factors respectively in the appendix(\ref{secttion.A}).

\section{Attenuation Characteristics Plots}

\begin{figure}
\centering
\includegraphics[width=1.0\textwidth]{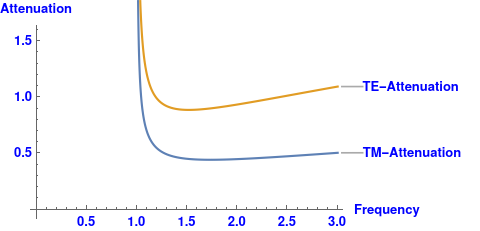}
\caption{A General Attenuation Constant Plot for $m=n=1$}
\label{GeneralACII}
\end{figure}

We give a general plot for the TM-mode and TE-mode attenuation constants of the ETW for $m=n=1$ in fig:\ref{GeneralACII} (plots for normalized attenuation and frequency). As can be seen from the plots, the curve in the lower region for the TE-mode, is far more abrupt or sharp compared to that of the TM-mode, this is a direct opposite of the situation with the elliptic waveguide\cite{GutierrezVegaIV}, where the TM modes are more abrupt.

\begin{figure}
\centering
\includegraphics[width=1.0\textwidth]{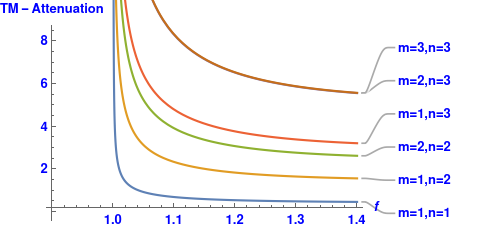}
\caption{General TM Attenuation}
\label{fig.Att_a}
\end{figure}

\begin{figure}
\centering
\includegraphics[width=1.0\textwidth]{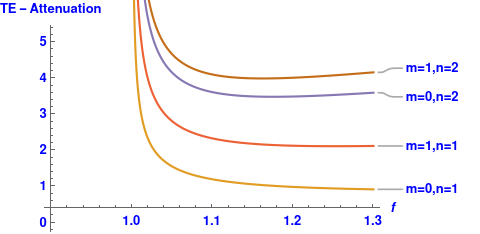}
\caption{General TE Attenuation}
\label{fig.Att_b}
\end{figure}

\begin{figure}
\centering
\includegraphics[width=1.0\textwidth]{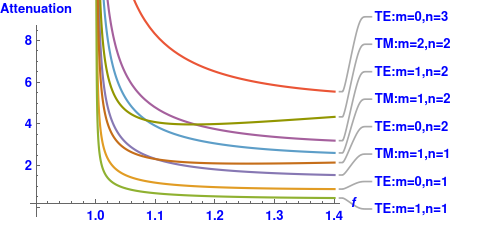}
\caption{Combined TM and TE attenuation Plots for the lowest modes}
\label{fig.Att_c}
\end{figure}

Now Figs:\ref{fig.Att_a} and \ref{fig.Att_b} illustrates the general attenuation curves (normalized attenuation constants against normalized frequencies) for both the TM-mode and TE-modes. Fig:\ref{fig.Att_c} shows a combined attenuation plots for the lowest modes. The TE-mode attenuation constant is more sensitive to the change in magnitude of the integers $l$,$m$ and $n$, than the TM-mode attenuation, because it has the factor $k_{\bot}$, to the second order of magnitude higher than that of the TM-mode as can be seen in Eqs.(\ref{attenuationIIa}) and (\ref{attenuationIIb}). For an attenuation constant plot, within the same range of attenuation, a little increment in the relative magnitudes of the indices $m$ and $n$ results in a considerable change and rise of the TE-mode curves as compared to the TM-mode curves. 

Now given that from the simple relation Eq:\ref{three.index.relation}, we have noted that all three indices m,n and l cannot be equal. So also, as is very well known, we cannot change the sign of any one (or two) of the indices without changing signs of the other two (or the remaining index)\cite{Milton,Borgnis}. We note in this respect, that the attenuation plots for all positive and all negative values of the same m and n indices, result in the two plots being exactly the same. Since, if by any means we changed the signs of m and n, then, the only meaningful outcome, is that we have also changed the sign of l simultaneously. Because $-l-m-n=0$ is exactly the same as Eq.\ref{three.index.relation}.


\section{Quality Factor}

\begin{figure}
\centering
\includegraphics[width=1.0\textwidth]{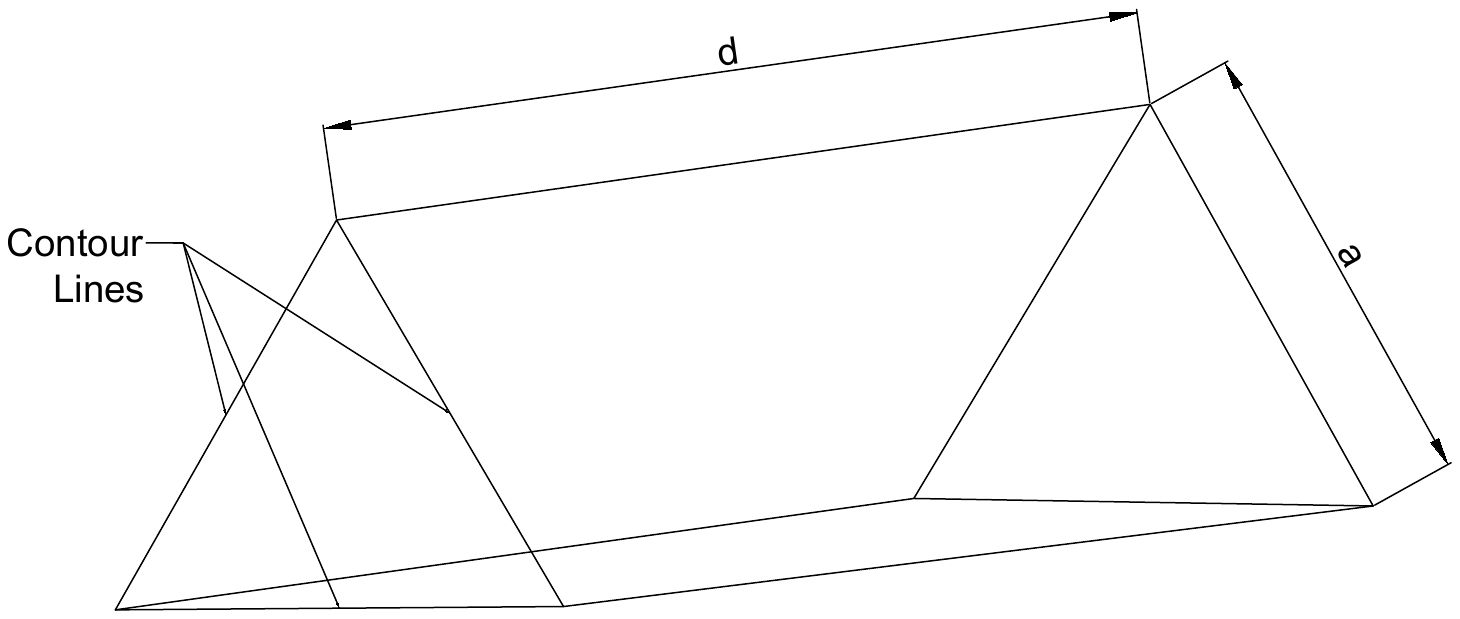}
\caption{A cross section of an ET Cavity}
\label{ETWCavity}
\end{figure}

If the plane boundary surfaces are at $z=0$ and $z=d$, then boundary conditions can be satisfied at each surface only if\cite{Jackson} $k=p\frac{\pi}{d}$. In which case the fields would be for TM fields
\begin{align}
&\mathbf{E}_{||}=\gamma_{p}^{2}\Phi\left(x,y\right)\cos\left(\frac{p\pi z}{d}\right)\quad{;}\quad H_{||}=0,\nonumber\\
&\mathbf{H}_{\bot}=i\omega\epsilon\cos\left(\frac{p\pi z}{d}\right)\hat{z}\times\mathbf{\nabla}_{\bot}\Phi\left(x,y\right),\nonumber\\
&\mathbf{E}_{\bot}=-\frac{p\pi}{d}\sin\left(\frac{p\pi z}{d}\right)\mathbf{\nabla}_{\bot}\Phi\left(x,y\right)\quad{;}\quad p=0,1,2,3,...,
\end{align}

and for the TE fields
\begin{align}
&\mathbf{H}_{||}=\gamma_{p}^{2}\Psi\left(x,y\right)\sin\left(\frac{p\pi z}{d}\right)\quad{;}\quad\mathbf{H}_{\bot}=\frac{p\pi}{d}\cos\left(\frac{p\pi z}{d}\right)\mathbf{\nabla}_{\bot}\Psi\left(x,y\right)\nonumber\\
&\mathbf{E}_{\bot}=-i\omega\mu\sin\left(\frac{p\pi z}{d}\right)\hat{z}\times\mathbf{\nabla}_{\bot}\Psi\left(x,y\right)\quad{;}\quad E_{||}=0,\quad p=1,2,3,...,
\end{align}

a sketch of the cavity is shown in fig:\ref{ETWCavity}.

The quality factor, is defined in general to be:
\begin{align}
    Q_{p}^{(TM)}=\omega_{p}\frac{U_{p}^{(TM)}}{P_{d}^{(TM)}}\quad{;}\quad Q_{p}^{(TE)}=\omega_{p}\frac{U_{p}^{(TE)}}{P_{d}^{(TE)}},
\end{align}

where
\begin{align}
    U_{p}=\frac{\mu}{2}\int_{V}\mathbf{H}_{p}\cdot\mathbf{H}_{p}^{*}dV,
\end{align}
\begin{align}
P_{d}=&\left[\int_{0}^{d}P_{L}^{\prime}dz+\frac{1}{4}\omega\mu\delta_{s}\left(\int_{S}\left(\mathbf{H}\cdot\mathbf{H}^{*}\right)da\vert_{z=0}\right.\right.\nonumber\\
&\left.\left.+\int_{S}\left(\mathbf{H}\cdot\mathbf{H}^{*}\right)da\vert_{z=d}\right)\right],
\end{align}

and $P_{L}^{\prime}=\frac{1}{4}\omega\mu\delta_{s}\oint_{C}\mathbf{H}\cdot\mathbf{H}^{*}dl$. With these we have in general for any cylindrical cavity, of length $d$ (wave number $k=p\frac{\pi}{d}$; $p=0,1,2,3,...$) with two covering ends perpendicular to the cylinder axis

\begin{align}
    Q_{p}^{(TM)}=\frac{d}{\left[\frac{\delta_{s}d}{2}\oint_{C}\vert\mathbf{\nabla}_{\bot}\Phi\left(x,y\right)\vert^{2}dl+\chi_{0p}\delta_{s}\right]},\quad\mathrm{and}
\label{TMQuality.factor}
\end{align}
\begin{align}
Q_{p}^{(TE)}=\frac{\gamma_{p}^{2}\left(\chi_{0p}-1\right)+2\frac{p^{2}\pi^{2}}{d^{2}\chi_{0p}}}{\left[\frac{\delta_{s}\gamma_{p}^{4}\left(\chi_{0p}-1\right)}{2d}\oint_{C}\vert\Psi\left(x,y\right)\vert^{2}dl+\frac{\delta_{s}p^{2}\pi^{2}}{d^{2}\chi_{0p}}\oint_{C}\vert\mathbf{\nabla}_{\bot}\Psi\left(x,y\right)\vert^{2}dl+2\delta_{s}\left(\frac{p^{2}\pi^{2}}{d^{3}}\right)\right]}.
\label{TEQuality.factor}
\end{align}

(With $\chi_{0p}=2-\delta_{0p}$ and $\delta_{0p}$ a Kronecker delta for p and 0).

As for the ETW, if we defined
\begin{align}
&I^{(TM)}=I_{1}^{(TM)}+I_{2}^{(TM)}+I_{3}^{(TM)};\nonumber\\&I_{i}^{(TM)}=\int_{c_{i}}\vert\mathbf{\nabla}_{\bot}\Phi_{lmn}\left(x,y\right)\vert^{2}dl_{i}\quad{;}\quad i=1,2,3.\quad\mathrm{And}\nonumber\\	&I_{a}^{(TE)}=I_{11}^{(TE)}+I_{12}^{(TE)}+I_{13}^{(TE)};\nonumber\\ &I_{b}^{(TE)}=I_{21}^{(TE)}+I_{22}^{(TE)}+I_{23}^{(TE)},\quad\mathrm{with:}\nonumber\\
&I_{1i}^{(TE)}=\int_{c_{i}}\vert\Psi_{lmn}\left(x,y\right)\vert^{2}dl_{i};\nonumber\\
&I_{2i}^{(TE)}=\int_{c_{i}}\vert\mathbf{\nabla}_{\bot}\Psi_{lmn}\left(x,y\right)\vert^{2}dl_{i}\quad{;}\quad i=1,2,3.
\end{align}

Then, this implies that
\begin{align}
&I^{(TM)}=\oint_{C}\vert\mathbf{\nabla}_{\bot}\Phi\left(x,y\right)\vert^{2}dl;\nonumber\\ &I_{a}^{(TE)}=\oint_{C}\vert\Psi\left(x,y\right)\vert^{2}dl;\nonumber\\ &I_{b}^{(TE)}=\oint_{C}\vert\mathbf{\nabla}_{\bot}\Psi\left(x,y\right)\vert^{2}dl.
\end{align}

Thus for the same ETW, the quality factors are
\begin{align}
Q_{p}^{(TM)}=\frac{d}{\left[\frac{\delta_{s}d}{2}I^{(TM)}+\chi_{0p}\delta_{s}\right]},
 \label{TMQuality.factorII}
\end{align}

and
\begin{align}
Q_{p}^{(TE)}=\frac{\gamma_{p}^{2}\left(\chi_{0p}-1\right)+2\frac{p^{2}\pi^{2}}{d^{2}\chi_{0p}}}{\left[\frac{\delta_{s}\gamma_{p}^{4}\left(\chi_{0p}-1\right)}{2d}I_{a}^{(TE)}+\frac{\delta_{s}p^{2}\pi^{2}}{d^{2}\chi_{0p}}I_{b}^{(TE)}+2\delta_{s}\left(\frac{p^{2}\pi^{2}}{d^{3}}\right)\right]},
\label{TEQuality.factorII}
\end{align}

with $\chi_{0p}=2-\delta_{0p}$. We have incorporated the integer "p" in the derivation, so that one would not have to keep track of whether or not it is the TM or TE fields, for which p is allowed to be zero. One can for instance confirm that for $p=0$, Eq.(\ref{TEQuality.factor}) or (\ref{TEQuality.factorII}) is undefined, which is in line with $p\neq0$ for the cavity TE fields. 

We could rewrite Eq.s(\ref{TMQuality.factorII}) and (\ref{TEQuality.factorII}) as
\begin{align}
&Q_{p}^{\prime\left(TM\right)}=\frac{ar}{\left[\frac{aI^{(TM)}}{2}r+\chi_{0p}\right]},\quad\mathrm{and}\nonumber\\
&Q_{p}^{\prime\left(TE\right)}=\frac{\gamma_{p}^{2}\left(\chi_{0p}-1\right)+2\frac{p^{2}\pi^{2}}{a^2r^{2}\chi_{0p}}}{\left[\frac{\gamma_{p}^4\left(\chi_{0p}-1\right)}{2ar}I_{a}^{(TE)}+\frac{p^{2}\pi^{2}}{a^{2}r^{2}\chi_{0p}}I_{b}^{(TE)}+2\frac{p^{2}\pi^{2}}{a^{3}r^{3}}\right]},
\label{TM.TE.Normalized.QF}
\end{align}

where

\begin{align}
&Q_{p}^{\prime(TM)}=Q_{p}^{(TM)}\delta_{s}\quad{;}\quad Q_{p}^{\prime(TE)}=Q_{p}^{(TE)}\delta_{s}\nonumber\\
&r=\frac{d}{a}.\quad\mathrm{Now}\quad\mathrm{if}\quad v=\frac{1}{r},
\end{align}
then we can also write
\begin{align}
&Q_{P}^{\prime(TM)}=\frac{1}{\left[\frac{1}{2}I^{(TM)}+\frac{\chi_{0p}}{a}v\right]};\nonumber\\
&Q_{p}^{\prime(TE)}=\frac{\gamma_{p}^{2}\left(\chi_{0p}-1\right)+2\frac{\left(p\pi v\right)^{2}}{\chi_{0p}a^{2}}}{\left[\frac{\gamma_{p}^{4}\left(\chi_{0p}-1\right)I_{a}^{(TE)}v}{2a}+\frac{p^{2}\pi^{2}I_{b}^{(TE)}}{a^{2}\chi_{0p}}v^{2}+2\frac{p^{2}\pi^{2}}{a^{3}}v^{3}\right]}.
\label{TM.TE.Normalized.QFII}
\end{align}

\begin{figure}
\centering
\includegraphics[width=1.0\textwidth]{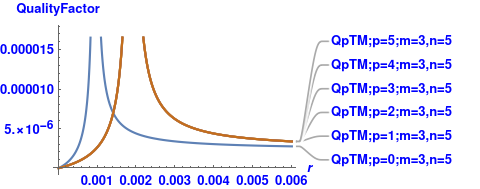}
\caption{TM Q factor for $p=0$, $p\neq0$ and fixed indices $m=3$,$n=5$}
\label{TMQa}
\end{figure}

\begin{figure}
\centering
\includegraphics[width=1.0\textwidth]{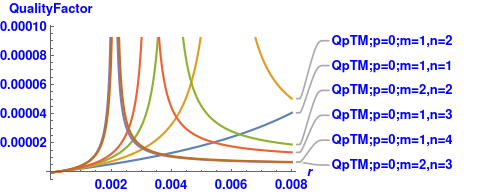}
\caption{ lowest modes TM Q factor plot for $p=0$}
 \label{TMQd}
\end{figure}

\begin{figure}
\centering
\includegraphics[width=1.0\textwidth]{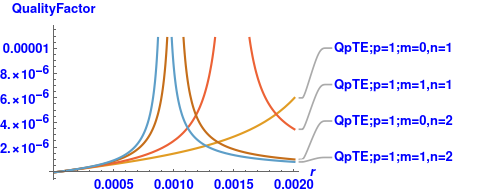}
\caption{lowest modes TE Q factor plot for $p=1$}
 \label{TEQe}
\end{figure}

\begin{figure}
\centering
\includegraphics[width=1.0\textwidth]{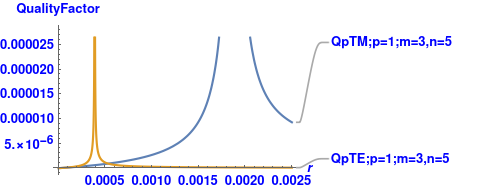}
\caption{TM and TE Q factors for $p=1$ and same m,n indices}
\label{TMTEQf}
\end{figure}

We note that a quality factor plot against v, produces the same result or plot as that against r. Referring to Eq.s(\ref{TM.TE.Normalized.QF}) and (\ref{TM.TE.Normalized.QFII})  as the normalized quality factors, one can see that, the TM quality factor is dependent on p, only through the Kronecker delta (see also Eq.s(\ref{TMQuality.factor}) and (\ref{TMQuality.factorII})). So that for $p\neq0$ and fixed m and n indices, the TM Q factor remains unchanged for any value of p. This is illustrated in fig:\ref{TMQa}, but changing the indices m and n results in different plots. The general TM Q factor for the lowest modes with $p=0$, is shown in fig:\ref{TMQd}, while the lowest modes TE Q factor plots for $p=1$, is shown in fig:\ref{TEQe}. Note that it is not possible to make a plot of several indices for $p=0$, like the one in fig:\ref{TMQd}, for the TE Q factors, since these would be undefined. The Q factor plots are just as they would be expected\cite{Jackson}. A comparison of the Q factor (similar to our attenuation constants plot in fig:\ref{GeneralACII}), for the TM and TE-mode, for the same integer $p=1$, and indices $m=3$,$n=5$ is shown in fig:\ref{TMTEQf}. For a numerical (FDTD,finite-difference time-domain technique) calculation of ETWs quality factor see \cite{YZHuangII,YZHuang}. Our quality factor plots can be compared with those given in the fig:5 of Huang et al\cite{YZHuangII}, which agrees with our own plots, even though the plots in \cite{YZHuangII} were a numerical plot of the q-factors against frequency.


We note that with
\begin{align}
    k_{\bot}^{2}=\frac{2\pi^{2}}{3h^{2}}\left(l^{2}+m^{2}+n^{2}\right)\quad{;}\quad h=\frac{\sqrt{3}}{2}a,
\end{align}

we have
\begin{align}
    k_{\bot}^{2}=\frac{8\pi^{2}}{9a^{2}}\left(l^{2}+m^{2}+n^{2}\right)=\frac{16\pi^{2}}{9a^{2}}\left(m^{2}+mn+n^{2}\right),
\end{align}

so that
\begin{align}	&\omega_{lmn}^{2}\mu\epsilon=k_{\bot}^{2}+\frac{p^{2}\pi^{2}}{d^{2}};\nonumber\\
&f_{lmn}=\frac{1}{2\pi\sqrt{\mu\epsilon}}\left(k_{\bot}^{2}+\frac{p^{2}\pi^{2}}{d^{2}}\right)\quad{;}\quad\lambda_{lmn}=\frac{2\pi}{\omega_{lmn}^{2}\mu\epsilon}.
\end{align}

\section{Conclusion}

We have derived general expressions for calculating the quality factors of any cylindrical wave guide, given their well known mode functions, from which we calculated the quality factors of the ETW. And using the analogous well known general expressions for the attenuation constants\cite{Borgnis}, we have found the attenuation constants of the ETW, also using their mode functions, by calculating some tedious but straight forward integrals of trigonometric functions, and have provided explicit solutions. 

We have plotted our attenuation constants, which are in agreement with the known results. We have also shown some special properties of the attenuation curves of the ETW. For instance, the abruptness or the sharpness of the TE and TM attenuation constant curves, at their turning points, which is in direct contrast to that of the elliptic waveguide curves\cite{GutierrezVegaIV} and have seen that, the larger the relative magnitudes of the (two) indices $m$,$n$ (say), the higher their attenuation constants. Also, the TE attenuation constants are more sensitive to the change in the relative magnitudes of the indices. 

In addition to the cyclic permutation property of the indices $l$, $m$ and $n$, we have also shown that the direct interchange (odd-permutation) of "m" and "n" (with the same "l"), is a symmetry transformation for the eigenfunctions, just as the change in the sign of "x" is, for the even and odd modal functions.

From our integrals, we have also calculated the quality factors of the ETW, using our general quality factor expressions. Hence, we have obtained a completely analytic, explicit and closed form solutions for the attenuation constants and quality factors of the ETW. 

We also presented a detailed table of the eigenvalues of the equilateral triangular waveguide, using the fundamental theorem of arithmetic and the theory of Eisenstein integers. Illustrating rigorously for the first time some subtle or accidental degeneracies in which either the odd modes or the TM modes are excluded in some mode sets but present in another mode sets of the same eigenvalues.  

We have shown various field plots and characteristics of the modal fields of the ETW, and have the seen the peculiar propagation properties of the fields when the two indices $m$ and $n$  are equal, in essence, the non-propagation of the odd modes, both for the TE and TM modes. Thus, we illustrated the special symmetric properties peculiar only to the (solutions of) ETWs. These, with our several plots of ETW attenuation constants and quality factors, show that the solutions of ETW are simple, with interesting properties revealed through the manipulation of the three indices $l$, $m$ and $n$. Thus making the ETW robust for various applications, especially in nano-photonics and/or nanotechnology. The modes of the ETW have found applications in billiards and our research will be invaluable for the further investigations of equilateral triangular billiards and other related areas, that studies light or particle interactions confined within the equilateral triangular geometry. 

Our analytic methods or scheme could be applied to other waveguides. The general expressions of the quality factors given here, just as that of the attenuation constants can also find application in other triangular (hemiequilateral, regular hexagonal and rhombus\cite{McCartinIII}) or polygonal cylindrical waveguides and the study of waveguides of arbitrary cross sections.

\section{Appendices}
\subsection{Transverse Modal Fields}
\label{TMTEtransverseFields}
\subsubsection{Transverse TM Fields}
The transverse TM fields are given by

\begin{align}
_{e}\mathbf{E}_{\bot}=2i\beta  \Biggl[&\mp\left(\theta_{1}\sin\left(\theta_{1}x\right)\sin\left(lQ\left(y\right)\right)+\theta_{2}\sin\left(\theta_{2}x\right)\sin\left(mQ\left(y\right)\right)+\theta_{3}\sin\left(\theta_{3}x\right)\sin\left(nQ\left(y\right)\right)\right)\hat{x}\Biggr.\nonumber\\
&\Biggl.\pm\frac{\pi}{h}\left(l\cos\left(\theta_{1}x\right)\cos\left(lQ\left(y\right)\right)+m\cos\left(\theta_{2}x\right)\cos\left(mQ\left(y\right)\right)+n\cos\left(\theta_{3}x\right)\cos\left(nQ\left(y\right)\right)\right)\hat{y}\Biggr]e^{\pm i\beta z-i\omega t},
\end{align}
\begin{align}
_{o}\mathbf{E}_{\bot}=\pm2i\beta  \Biggl[&\left(\theta_{1}\cos\left(\theta_{1}x\right)\sin\left(lQ\left(y\right)\right)+\theta_{2}\cos\left(\theta_{2}x\right)\sin\left(mQ\left(y\right)\right)+\theta_{3}\cos\left(\theta_{3}x\right)\sin\left(nQ\left(y\right)\right)\right)\hat{x}\Biggr.\nonumber\\
&\Biggl.+\frac{\pi}{h}\left(l\sin\left(\theta_{1}x\right)\cos\left(lQ\left(y\right)\right)+m\sin\left(\theta_{2}x\right)\cos\left(mQ\left(y\right)\right)+n\sin\left(\theta_{3}x\right)\cos\left(nQ\left(y\right)\right)\right)\hat{y}\Biggr] e^{\pm i\beta z-i\omega t}.
\end{align}

With
\begin{align}
    _{\xi}\mathbf{H}_{\bot}=\pm\frac{1}{Z_{1}}\left(\hat{z}\times\left(_{\xi}\mathbf{E}_{\bot}\right)\right)\quad{;}\quad Z_{1}=\frac{\beta}{\omega\epsilon},
\end{align}
\subsubsection{Transverse TE Fields}
The transverse TE fields are given by

\begin{align}
_{e}\mathbf{H}_{\bot}=\mp2i\beta\Biggl[&\left(\theta_{1}\sin\left(\theta_{1}x\right)\cos\left(lQ\left(y\right)\right)+\theta_{2}\sin\left(\theta_{2}x\right)\cos\left(mQ\left(y\right)\right)+\theta_{3}\sin\left(\theta_{3}x\right)\cos\left(nQ\left(y\right)\right)\right)\hat{x}\Biggr.\nonumber\\
&\Biggl.+\frac{\pi}{h}\left(l\cos\left(\theta_{1}x\right)\sin\left(lQ\left(y\right)\right)+m\cos\left(\theta_{2}x\right)\sin\left(mQ\left(y\right)\right)+n\cos\left(\theta_{3}x\right)\sin\left(nQ\left(y\right)\right)\right)\hat{y}\Biggr] e^{\pm i\beta z-i\omega t},
\end{align}
\begin{align}
_{o}\mathbf{H}_{\bot}=2i\beta\Biggl[&\pm\left(\theta_{1}\cos\left(\theta_{1}x\right)\cos\left(lQ\left(y\right)\right)+\theta_{2}\cos\left(\theta_{2}x\right)\cos\left(mQ\left(y\right)\right)+\theta_{3}\cos\left(\theta_{3}x\right)\cos\left(nQ\left(y\right)\right)\right)\hat{x}\Biggr.\nonumber\\
&\Biggl.\mp\frac{\pi}{h}\left(l\sin\left(\theta_{1}x\right)\sin\left(lQ\left(y\right)\right)+m\sin\left(\theta_{2}x\right)\sin\left(mQ\left(y\right)\right)+n\sin\left(\theta_{3}x\right)\sin\left(nQ\left(y\right)\right)\right)\hat{y}\Biggr]e^{\pm i\beta z-i\omega t},
\end{align}

with
\begin{align}
    _{\xi}\mathbf{E}_{\bot}=\pm Z_{2}\hat{z}\times\left(_{\xi}\mathbf{H}_{\bot}\right)\quad{;}\quad Z_{2}=\frac{\omega\mu}{\beta}.
\end{align}
\subsection{Integral Solutions and Parameters for Attenuation Constants and Quality Factors}
\label{secttion.A}
On computing the integrals we obtain
\begin{align}
I_{1}^{(TM)}=&\frac{4\pi^{2}\gamma^{\prime 2}_{lmn}a}{h^{2}}+16\left(\frac{lm}{n}\right)\frac{\pi}{\sqrt{3}h}(-1)^{l+m}\sin(a_{n})\nonumber\\
&+16\left(\frac{ln}{m}\right)\frac{\pi}{\sqrt{3}h}\left(-1\right)^{l+n}\sin\left(a_{m}\right)\nonumber\\
 &+16\left(\frac{mn}{l}\right)\frac{\pi}{\sqrt{3}h}\left(-1\right)^{m+n}\sin\left(a_{l}\right);
\end{align}
\begin{align}
I_{11}^{(TE)}=&12a+\frac{16h\left(-1\right)^{l+m}}{\sqrt{3}n\pi}\sin\left(a_{n}\right)\nonumber\\
 &+\frac{16h\left(-1\right)^{n+l}}{\sqrt{3}m\pi}\sin\left(a_{m}\right)\nonumber\\
 &+\frac{16h\left(-1\right)^{m+n}}{\sqrt{3}l\pi}\sin\left(a_{l}\right);
\end{align}
\begin{align}
I_{21}^{(TE)}=&4a\Theta+\frac{16h\theta_{1}\theta_{2}\left(-1\right)^{l+m}}{\sqrt{3}n\pi}\sin\left(a_{n}\right)\nonumber\\
 &+\frac{16h\theta_{1}\theta_{3}\left(-1\right)^{l+n}}{\sqrt{3}m\pi}\sin\left(a_{m}\right)\nonumber\\
 &+\frac{16h\theta_{2}\theta_{3}\left(-1\right)^{m+n}}{\sqrt{3}l\pi}\sin\left(a_{l}\right),\quad\mathrm{and}
\end{align}

\begin{align}
I_{2}^{(TM)}=I_{3}^{(TM)}=-4t_{1}\Biggl[&\Delta_{lmn}+p_{1}\left(g_{12}-g_{11}\right)+p_{2}\left(g_{22}-g_{21}\right)+p_{3}\left(g_{32}-g_{31}\right)\Biggr.\nonumber\\	
&\Biggl.+z_{1}\left(g_{11}+g_{12}\right)+z_{2}\left(g_{21}+g_{22}\right)+z_{3}\left(g_{31}+g_{32}\right)\Biggr];
\end{align}

\begin{align}
I_{12}^{(TE)}=I_{13}^{(TE)}=-2t_{1}\Biggl[\frac{3a}{2}+g_{11}+g_{12}+g_{21}+g_{22}+g_{31}+g_{32}\Biggr].\quad\mathrm{While},
\end{align}

\begin{align}
I_{22}^{(TE)}=I_{23}^{(TE)}=-4t_{1}\Biggl[&\Delta_{lmn}+p_{1}\left(g_{11}+g_{12}\right)+p_{2}\left(g_{21}+g_{22}\right)+p_{3}\left(g_{31}+g_{32}\right)\Biggr.\nonumber\\	
&\Biggl.+z_{1}\left(g_{12}-g_{11}\right)+z_{2}\left(g_{22}-g_{21}\right)+z_{3}\left(g_{32}-g_{31}\right)\Biggr].
\end{align}

Where
\begin{align}	&\Theta=\theta_{1}^{2}+\theta_{2}^{2}+\theta_{3}^{2}=\left(1+\frac{3a^{2}}{8h^{2}}\right)k_{\bot}^{2}\quad{;}\quad\gamma^{\prime 2}_{lmn}=l^{2}+m^{2}+n^{2}=\frac{3h^{2}k_{\bot}^{2}}{2\pi^{2}}\quad{;}\quad l+m+n=0.\nonumber\\
&a_{l}=\frac{\sqrt{3}al}{2h}\pi\quad{;}\quad a_{m}=\frac{\sqrt{3}am}{2h}\pi\quad{;}\quad a_{n}=\frac{\sqrt{3}an}{2h}\pi\quad{;}\quad \Delta_{lmn}=\frac{3k_{\bot}^{2}a}{4}+\frac{a}{4}\left(F_{l}+F_{m}+F_{n}\right);\nonumber\\
&F_{l}=\theta_{1}^{2}-\frac{l^{2}\pi^{2}}{h^{2}}\quad{;}\quad F_{m}=\theta_{2}^{2}-\frac{m^{2}\pi^{2}}{h^{2}}\quad{;}\quad F_{n}=\theta_{3}^{2}-\frac{n^{2}\pi^{2}}{h^{2}}\quad{;}\quad k_{\bot}^{2}=\frac{2\pi^{2}}{3h^{2}}\left(l^{2}+m^{2}+n^{2}\right);\nonumber\\
&\theta_{1}=\frac{\pi}{\sqrt{3}h}\left(m-n\right)\quad{;}\quad\theta_{2}=\frac{\pi}{\sqrt{3}h}\left(n-l\right)\quad{;}\quad\theta_{3}=\frac{\pi}{\sqrt{3}h}\left(l-m\right);\quad{;}\quad\beta=k^{2}_{z}=\omega^{2}\mu\epsilon-k_{\bot}^{2}\nonumber\\
&p_{1}=\frac{\theta_{1}\theta_{2}}{2}\quad{;}\quad p_{2}=\frac{\theta_{1}\theta_{3}}{2}\quad{;}\quad p_{3}=\frac{\theta_{2}\theta_{3}}{2}\quad{;}\quad z_{1}=\frac{lm\pi^{2}}{2h^{2}}\quad{;}\quad z_{2}=\frac{ln\pi^{2}}{2h^{2}}\quad{;}\quad z_{3}=\frac{mn\pi^{2}}{2h^{2}};\nonumber\\
&\delta_{s}=\sqrt{\frac{2}{\omega\mu_{2c}\sigma_{2c}}}\quad{;}\quad h=\frac{a}{2}\sqrt{3}\quad{;}\quad t_{1}=\frac{1}{a}\left(a^{2}+4h^{2}\right)^{\frac{1}{2}},\nonumber\\
\end{align}

\begin{align}
&q_{11}=\frac{\sqrt{3}an\pi+2h\pi\left(l+m\right)}{ah}\quad{;}\quad q_{12}=\frac{\sqrt{3}na\pi-2h\pi\left(l+m\right)}{ah}\quad{;}\quad q_{13}=\frac{\sqrt{3}an\pi+2h\pi\left(l-m\right)}{ah};\nonumber\\
&q_{14}=\frac{\sqrt{3}an\pi-2\pi h\left(l-m\right)}{ah}\quad{;}\quad q_{21}=\frac{\sqrt{3}ma\pi-2\pi h\left(l+n\right)}{ah}\quad{;}\quad q_{22}=\frac{\sqrt{3}ma\pi+2\pi h\left(l+n\right)}{ah};\nonumber\\
&q_{23}=\frac{\sqrt{3}ma\pi-2\pi h\left(l-n\right)}{ah}\quad{;}\quad q_{24}=\frac{\sqrt{3}ma\pi+2\pi h\left(l-n\right)}{ah}\quad{;}\quad q_{31}=\frac{\sqrt{3}la\pi+2\pi h\left(m+n\right)}{ah};\nonumber\\
&q_{32}=\frac{\sqrt{3}la\pi-2\pi h\left(m+n\right)}{ah}\quad{;}\quad q_{33}=\frac{\sqrt{3}la\pi+2\pi h\left(m-n\right)}{ah}\quad{;}\quad q_{34}=\frac{\sqrt{3}la\pi-2\pi h\left(m-n\right)}{ah};\nonumber\\
&\mu_{11}=\frac{q_{11}a}{2}\quad{;}\quad\mu_{12}=\frac{q_{12}a}{2}\quad{;}\quad\mu_{13}=\frac{q_{13}a}{2}\quad{;}\quad\mu_{14}=\frac{q_{14}a}{2};\nonumber\\
&\mu_{21}=\frac{q_{21}a}{2}\quad{;}\quad\mu_{22}=\frac{q_{22}a}{2}\quad{;}\quad\mu_{23}=\frac{q_{23}a}{2}\quad{;}\quad\mu_{24}=\frac{q_{24}a}{2};\nonumber\\
&\mu_{31}=\frac{q_{31}a}{2}\quad{;}\quad\mu_{32}=\frac{q_{32}a}{2}\quad{;}\quad\mu_{33}=\frac{q_{33}a}{2}\quad{;}\quad\mu_{34}=\frac{q_{34}a}{2},\quad\mathrm{with}
\end{align}

\begin{align}
&g_{11}=\frac{1}{q_{11}}\sin\left(\mu_{11}\right)+\frac{1}{q_{12}}\sin\left(\mu_{12}\right)\quad{;}\quad g_{12}=\frac{1}{q_{13}}\sin\left(\mu_{13}\right)+\frac{1}{q_{14}}\sin\left(\mu_{14}\right);\nonumber\\
&g_{21}=\frac{1}{q_{21}}\sin\left(\mu_{21}\right)+\frac{1}{q_{22}}\sin\left(\mu_{22}\right)\quad{;}\quad g_{22}=\frac{1}{q_{23}}\sin\left(\mu_{23}\right)+\frac{1}{q_{24}}\sin\left(\mu_{24}\right);\nonumber\\
&g_{31}=\frac{1}{q_{31}}\sin\left(\mu_{31}\right)+\frac{1}{q_{32}}\sin\left(\mu_{32}\right)\quad{;}\quad g_{32}=\frac{1}{q_{33}}\sin\left(\mu_{33}\right)+\frac{1}{q_{34}}\sin\left(\mu_{34}\right).
\end{align}

\begin{acknowledgments}
F.~E.~O. acknowledges financial support from CONACyT (Consejo Nacional de Ciencia y Tecnolog\'ia) and Tec de Monterrey (ITESM). He also thanks U.N.N. for study leave support. F.E.O. also wants to acknowledge his lovely mother, Regina Ngozi Onah, who died on the 7th of March 2025. He dedicates his contribution to this work to her memory. J.~C.~G-V. acknowledges Tec de Monterrey (ITESM) and CONACyT for their financial support.
\end{acknowledgments}
\bibliography{references}


\end{document}